\documentclass[traditabstract]{aa}
\usepackage{graphicx}
\usepackage[varg]{txfonts}
\usepackage{natbib}
\bibpunct{(}{)}{;}{a}{}{,} 
\newcommand\ulsr{U_{\rm LSR}}
\newcommand\vlsr{V_{\rm LSR}}
\newcommand\wlsr{W_{\rm LSR}}

\newcommand\teff{T_{\rm eff}}

\begin{document}


\title{Exploring the production and depletion of lithium\\ in the Milky Way stellar disk\thanks{This paper includes data gathered with the 6.5\,m Magellan
Telescopes located at the Las Campanas Observatory, Chile;
the Nordic Optical Telescope (NOT) on La Palma, Spain;
the Very Large Telescope (VLT) at the European Southern Observatory
(ESO) on Paranal, Chile (ESO Proposal ID 69.B-0277 and 72.B-0179);
the ESO\,1.5\,m, 2.2\,m, and 3.6\,m telescopes on
La Silla, Chile (ESO Proposal ID 65.L-0019, 67.B-0108, 76.B-0416, 82.B-0610);
and data from the UVES Paranal Observatory Project
(ESO DDT Program ID 266.D-5655).}
\fnmsep
\thanks{
Table~\ref{tab:parameters} is only available in electronic form at the CDS via anonymous ftp to
\texttt{cdsarc.u-strasbg.fr (130.79.128.5)} or via
\texttt{http://cdsweb.u-strasbg.fr/cgi-bin/qcat?J/A+A/XXX/AXX}.
}}
\titlerunning{}
\author{
Thomas Bensby\inst{1}
\and
Karin Lind\inst{2,3}
}

\institute{
Lund Observatory, Department of Astronomy and Theoretical physics,
Box 43, SE-221\,00 Lund, Sweden\\
\email{tbensby@astro.lu.se}
\and
Max Planck Institute for Astronomy, K\"onigstuhl 17, 69117 Heidelberg, Germany
\and
Department of Physics and Astronomy, Uppsala University, Box 516, SE-751\,20 Uppsala, Sweden
}

\date{Received 28 March 2018 / Accepted 19 April 2018}
\abstract{
Despite the recent availability of large samples of stars with high-precision Li abundances, there are many unanswered questions about the evolution of this unique element in the Galaxy and in the stars themselves.
It is unclear which parameters and physical mechanisms that govern Li depletion in late-type stars and if Galactic enrichment has proceeded differently in different stellar populations. With this study we aim to explore these questions further by mapping the evolution of Li with stellar mass, age, and effective temperature for Milky Way disk stars, linking the metal-poor and metal-rich regimes, and how Li differs in the thin and thick disks. We determine Li abundances for a well-studied sample of 714 F and G dwarf, turn-off, and subgiant stars in the solar neighbourhood. The analysis is based on line synthesis of the $^7$Li line at 6707\,{\AA} in high-resolution and high signal-to-noise ratio echelle spectra, obtained with the MIKE, FEROS, SOFIN, UVES, and FIES spectrographs. The presented Li abundances are corrected for non-LTE effects. Out of the sample of 714 stars we are able to determine Li abundances for 420 stars and upper limits on the Li abundance for another 121 stars. 36 stars are listed as exoplanet host stars, and 18 of those have well-determined Li abundances and 6 have Li upper limits. Our main finding is that there are no signatures of Li production in stars associated with the thick disk. Instead the Li abundance trend is decreasing with metallicity for these thick disk stars. Significant Li production is however seen in the thin disk, with a steady increase towards super-solar metallicities. At the highest metallicities, however, around $\rm [Fe/H]\approx +0.3$, we tentatively confirm the recent discovery that the Li abundances level out. Our finding contradicts the other recent studies that found that Li is also produced in the thick disk. We find that this is likely due to the $\alpha$-enhancement criteria those studies used to define their thick disk samples. By using the more robust age criteria we are able to define a thick disk stellar sample that is much less contaminated by thin disk stars. Furthermore, we also tentatively confirm the age-Li correlation for solar twin stars, and we find that there is no correlation between Li abundance and whether the stars have detected exoplanets or not. The major conclusion that can be drawn from this study is that no significant Li production, relative to the primordial abundance, took place during the first few billion years of the Milky Way, an era coinciding with the formation and evolution of the thick disk. Significant Li enrichment then took place once long-lived low-mass stars (acting on a time-scale longer than SNIa) have had time to contribute to the chemical enrichment of the interstellar medium.
}

\maketitle

\section{Introduction}

It has for a long time been known that the surface abundance of Li in solar-type stars changes as they evolve from the pre-main sequence, through the main sequence, to the evolved phases on the red giant branch \cite[e.g][]{bodenheimer1965,wallerstein1969,cayrel1984,michaud1991}. The most obvious example of Li depletion at work is the Sun whose photospheric abundance has decreased by more than two orders of magnitude since the birth of the solar system \citep[e.g.][]{asplund2009}. For main sequence and turn-off stars in general, Li depletion manifests itself observationally in two prominent features; the `Li dip' in open clusters, first discovered in the Hyades \citep{boesgaard1986}, and the `Spite Plateau'  of warm, metal-poor stars \citep{spite1982}. Both features reflect on the strong link between the efficiency of Li depletion and the stellar effective temperature; stars in a small $T_{\rm eff}$ interval between approximately $6000-6400$\,K are most stable against Li depletion, while both cooler and hotter stars significantly deplete their birth Li over time \citep{sestito2005}. Observationally, Li is seen to increase with increasing effective temperature from the bottom of the main sequence until the warm star plateau. For the hotter main sequence, which only exists for Pop I stars, a dip-like feature appears in A(Li) vs. $\teff$, centred around 6700\,K \citep{boesgaard2016}. 

The complex pattern of Li abundances observed in different types of stars cannot be explained with canonical stellar models, in particular because the bottom of the convection zone is not hot enough to destroy Li for roughly solar-mass main-sequence stars \citep[for a review see][]{pinsonneault1997}. Additional physical processes are therefore needed to consistently account for the transport of elements in the stellar interiors. From theoretical arguments, the most important properties for Li surface depletion are the temperature and mass at the bottom of the convection zone, as well as the efficiency of mixing in the radiative zone beneath \citep{richard2005}. Since the former is directly linked to the effective temperature on the main sequence, one may expect a simple, continuous Li-$\teff$ dependence for stars of a given age, contrary to what is observed. \citet[][]{talon2005} showed that the key to understanding the formation of the Li dip in open clusters and the inefficient Li depletion of stars on the plateau is the inclusion of atomic diffusion, rotational mixing, meridional circulation, and gravity waves in stellar models.

Both observations and theoretical predictions of the Li-$\teff$-relation agree that it is strongly age-related the first hundreds of millions of years, meaning that the Li abundance of stars decrease with time, with varying speed depending on the stellar mass, or equivalently for a single metallicity, the effective temperature.  However, while there is a possible indication for a halt in Li depletion after about two billion years in some open clusters \citep{sestito2005}, the stellar models predict an un-interrupted steady decrease for at least five to six billion years at solar metallicities \citep{deliyannis1997,charbonnel2005}. This is in agreement with the solar twin work pointing to a continuous depletion of A(Li) with age \citep[e.g.][]{carlos2016}.

In addition to the temporal evolution, the dependence of Li abundances on stellar mass (or effective temperature) and metallicity has been widely debated \citep[e.g.][]{asplund2006,bonifacio2007,aoki2009}. Most famously, the fine structure of the Spite Plateau of metal-poor stars has been investigated in the context of the missing cosmological Li problem, that is, the offset of about a factor of five between the observed abundances and the prediction from standard Big Bang nucleosynthesis  \citep[e.g.][]{melendez2010li,sbordone2010}. As suggested by several authors, stellar models with atomic diffusion and non-standard mixing may account for some or all of the missing Li and thus provide a solution to the cosmological dilemma \citep[e.g.][]{deliyannis1991,salaris2001,richard2005,korn2006}.  However, some studies have also claimed that Li depletion cannot be a simple function of age, mass, and metallicity, but must depend on yet more parameters, such as chromospheric activity, rotational velocity, binarity, and/or the presence of planets \citep[e.g.][]{ryan2002,strassmeier2012}. This might explain for example why main sequence stars in open cluster M67 still show a significant scatter in Li abundance at a given mass \citep{pace2012}.

While the Li variations with stellar mass and stellar age of stars of all evolutionary phases can be studied in open clusters, the observations are limited to high metallicity; the most metal-poor cluster with Li measurements is around $\rm [Fe/H]\approx-0.5$ \citep[e.g.][]{francois2013}. On the other hand, observations of Li in metal-poor globular clusters have so far only been able to reach stars at turn-off region in the HR diagram \citep[e.g.][]{pasquini2005,lind2009b,gonzalezhernandez2009,monaco2010,monaco2012,dobrovolskas2014}. 
Samples of field stars in the nearby Galactic disk and halo field may bridge the low and high metallicity regimes. Recent studies aimed at doing that \citep[e.g.][]{ramirez2012,delgadomena2015,guiglion2016,fu2018} agree that the upper envelope of Galactic Li abundance increases above the primordial Spite plateau level at higher metallicities. This is mainly attributed to Li production in the thin disk. The evolution of Li in the thick disk is less clear, whether it is increasing, decreasing, or remains flat.

In this work we will analyse 714 nearby F and G dwarf and subgiant stars in the Galactic disk from \cite{bensby2014}, simultaneously investigating the link between stellar age, mass, metallicity, and Li depletion, with special attention on differences in Li between the thin and thick disks.

\begin{table}[b]
\centering
\caption{Line list and atomic data\tablefootmark{$\dagger$}.
\label{tab:linelist}
}
\setlength{\tabcolsep}{1.55mm}
\footnotesize
\begin{tabular}{ccccccc}
\hline\hline
\noalign{\smallskip}
     & Wavelength &  EP &          & \multicolumn{3}{c}{Damping parameters} \\
Elm  & [\AA]      &  [eV] & $\log(gf)$ & Rad.&  Stark   & Waals    \\
\noalign{\smallskip}
\hline
\noalign{\smallskip}
\ion{Fe}{ii}& 6706.885 & 5.956 & $-$4.103 & 8.580 & $-$6.610 & $-$7.814\\
\ion{Si}{i} & 6706.979 & 5.954 & $-$2.560 & 8.150 & $-$3.170 & $-$6.930\\
\ion{Fe}{i} & 6707.172 & 5.538 & $-$2.810 & 8.150 & $-$3.440 & $-$7.120\\
\ion{Fe}{i} & 6707.431 & 4.608 & $-$2.250 & 8.300 & $-$4.470 & $-$7.480\\
\ion{V}{i}  & 6707.518 & 2.743 & $-$0.395 & 7.169 & $-$6.043 & $-$7.839 \\
\ion{Cr}{i} & 6707.596 & 4.207 & $-$2.667 & 7.170 & $-$5.770 & $-$7.790\\
\ion{Li}{i} & 6707.756 & 0.000 & $-$0.427 & 7.560 & $-$5.780 & $-$7.574 \\
\ion{Li}{i} & 6707.768 & 0.000 & $-$0.206 & 7.560 & $-$5.780 & $-$7.574 \\
\ion{Li}{i} & 6707.907 & 0.000 & $-$0.932 & 7.560 & $-$5.780 & $-$7.574 \\
\ion{Li}{i} & 6707.908 & 0.000 & $-$1.161 & 7.560 & $-$5.780 & $-$7.574 \\
\ion{Li}{i} & 6707.918 & 0.000 & $-$0.712 & 7.560 & $-$5.780 & $-$7.574 \\
\ion{Li}{i} & 6707.920 & 0.000 & $-$0.932 & 7.560 & $-$5.780 & $-$7.574 \\
\ion{V}{i}  & 6708.110 & 1.218 & $-$2.922 & 7.600 & $-$6.140 & $-$7.780\\
\ion{Fe}{i} & 6708.282 & 4.988 & $-$2.700 & 8.670 & $-$4.960 & $-$7.310\\
\ion{Fe}{i} & 6708.348 & 5.485 & $-$2.580 & 7.960 & $-$3.640 & $-$7.130\\
\ion{Fe}{i} & 6708.535 & 5.558 & $-$2.936 & 8.300 & $-$4.400 & $-$7.120\\
\ion{Fe}{i} & 6708.577 & 5.446 & $-$2.684 & 8.490 & $-$5.300 & $-$7.400\\
\noalign{\smallskip}
\hline
\end{tabular}
\tablefoot{
\tablefoottext{$\dagger$}{
Column~1 gives the element and degree of ionisation, col.~2 the wavelength, col.~3 the lower excitation potential, col.~4 the oscillator strength, and cols.~5--7 the damping constants. The $^7$Li components and nearby atomic lines are listed. Note that many linelists include a \ion{Si}{i} line 6708.023\,{\AA}. That inclusion dates back to \cite{mandell2004} that tried to fill an un-identified absorption feature assuming it is a highly excited \ion{Si}{i} line. However, new laboratory measurements of Si show no sign whatsoever of a Si line at that wavelength (Henrik Hartmann, private communication), and is not included.
}
}
\end{table}

\begin{figure*}
\centering
\resizebox{0.85\hsize}{!}{
        \includegraphics[angle=-90]{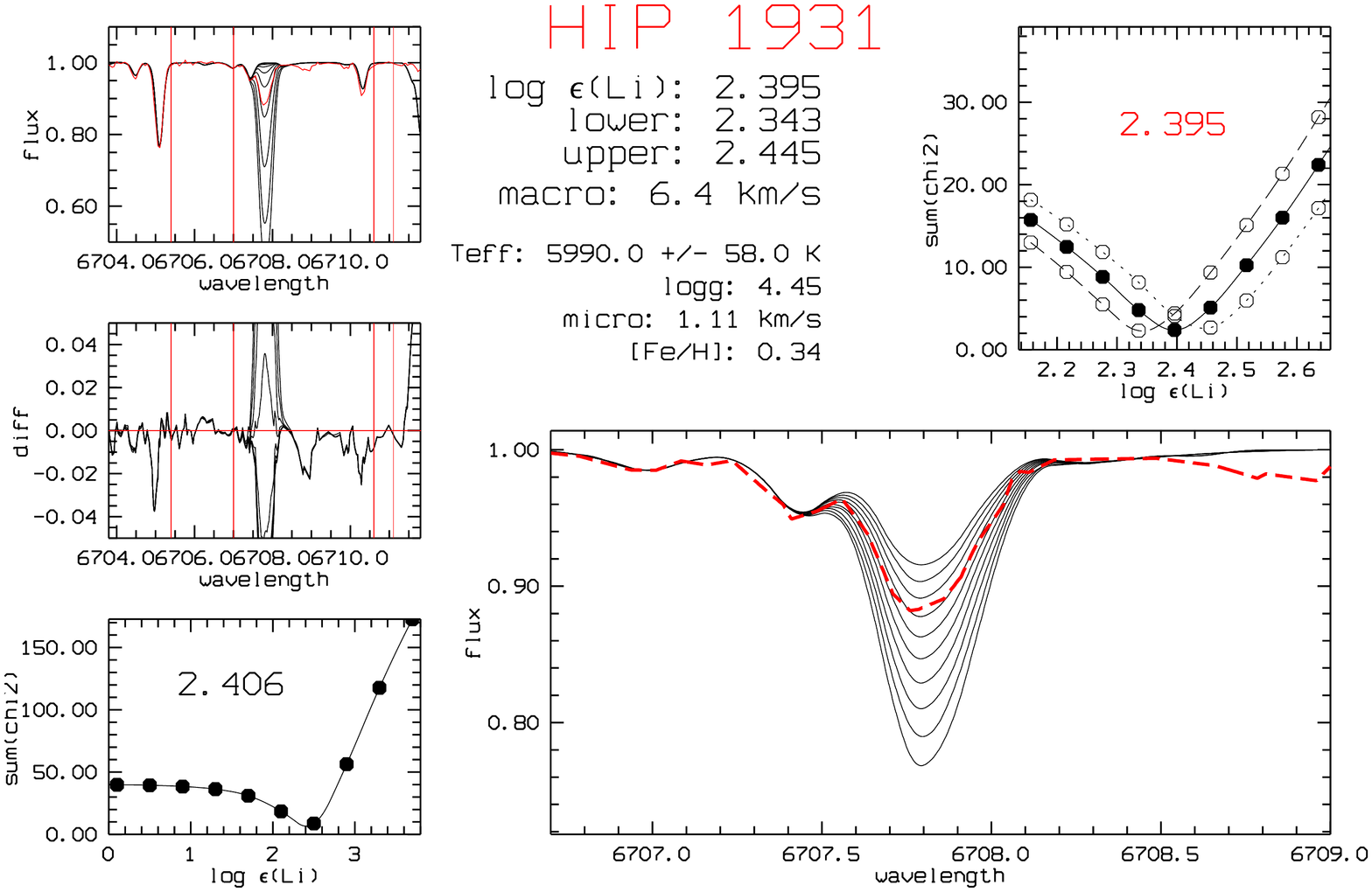}}
\caption{The fitting routine for the Li feature at 670.7\,nm. A grid of spectra with 10 different Li abundances in steps of 0.3\,dex is created in the range 0.1 to 3.7\,dex to get a first estimate. This is shown in the three panels on the left-hand side: {\sl top} the observed spectrum (red line) and 10 synthetic spectra; {\sl middle} the difference between observed and synthetic spectra; {\sl bottom} the sum of the squared difference as a function of Li abundance, where the minimum gives a first estimate of the Li abundance. A new set of 10 spectra with Li in steps of 0.06\,dex around the first estimate is then created and shown in the large panel on the right-hand side. The sum of the squared difference between observed and synthetic spectra is used to find the best abundance (shown in the top panel on right-hand side). The process is repeated taking the uncertainties in the effective temperature into account, giving the upper and lower uncertainties of the Li abundance (shown by the empty circles in the top panel on the right-hand side).
}
\label{fig:hip1931}
\end{figure*}

\section{Sample and analysis}

\subsection{Stellar sample and fundamental parameters}

The stellar sample consists of 714 F and G dwarf and subgiant stars from \cite{bensby2014}, aimed at exploring the age and abundance structure of the Galactic disk in the solar neighbourhood. The reader is directed to that paper for a full description of the observations, data reductions, determination of stellar parameters, stellar ages, stellar masses, as well as detailed abundances for 13 elements (O, Na, Mg, Al, Si, Ca, Ti, Cr, Fe, Ni, Zn, Y and Ba).  Elemental abundances for odd iron-peak elements (Sc, V, Mn, and Co) for the sample are presented in \cite{battistini2015}, and for neutron-capture elements (Sr, Zr, La, Ce, Nd, Sm, and Eu) in \cite{battistini2016}. Here we only give a brief summary of the observations and stellar parameter determination.

First, all stars were observed with the high-resolution spectrographs (FEROS -- \citealt{kaufer1999}; UVES -- \citealt{dekker2000}; MIKE -- \citealt{bernstein2003}; SOFIN -- \citealt{ilyin2000}; HARPS -- \citealt{mayor2003}; and FIES on the Nordic Optical Telescope) giving spectra with $R=45\,000$ to 110\,000 and signal-to-noise ratios over 200. The determination of stellar parameters and elemental abundances was based on equivalent width measurements and one-dimensional, plane-parallel, LTE model stellar atmospheres calculated with the Uppsala MARCS code \citep{gustafsson1975,edvardsson1993,asplund1997}.  The effective temperature $(\teff$) was determined by requiring excitation balance of abundances from Fe\,{\sc i} lines, the surface gravity ($\log g$) by requiring ionisation balance between abundances from Fe\,{\sc i} and Fe\,{\sc ii} lines. The microturbulence parameter ($\xi_{\rm t}$) was obtained by requiring that abundances from Fe\,{\sc i} lines are independent of line strength. In every step of the analysis NLTE corrections from \cite{lind2012} were applied to the abundances from individual Fe\,{\sc i} lines. 

Stellar ages were determined from a fine grid of $\alpha$-enhanced Yonsei-Yale (Y2) isochrones by \cite{demarque2004}, adopting $\rm [\alpha/Fe] = 0$ for $\rm [Fe/H] > 0$, $\rm [\alpha/Fe] = -0.3 \times [Fe/H]$ for $\rm -1 \leq [Fe/H] \leq 0$, and $\rm [\alpha/Fe] = +0.3$ for $\rm [Fe/H] < -1$. Taking the errors in effective temperature, surface gravity, and metallicity into account, an age probability distribution (APD) was constructed for each star. The most likely age, as well as lower and upper age estimates, was estimated from this APD as described in \cite{melendez2012}. In a similar manner, stellar masses were determined as well. As shown in \cite{bensby2017} this method gives very similar ages to those that can be estimated from more sophisticated Bayesian methods such as the one by \cite{jorgensen2005}.

\subsection{Synthesis of the Li feature at 670.7 nm}

In solar-type stars the only good indicator of Li is the resonance line at 670.7\,nm.  An option could be to use the Li line at 601.4\,nm, which, however, is too weak to be usable in normal dwarf stars (typically 10 to 100 times weaker than the 670.7\,nm line). Due to a number of features a single Gaussian profile is not a good approximation to the $^7$Li 670.7\,nm line profile. The line has fine structure splitting of about 0.15\,{\AA}, and on an even smaller scale hyperfine structure splitting, which in practise is too small to be influential.  $^6$Li might also be present and contribute to the Li feature.  In principle one could assume a $^6$Li/$^7$Li ratio increasing from 0\,\% at low metallicity \citep[SBBN does not produce $^6$Li, see for example][]{cyburt2016} to about 8\,\% at solar metallicity \citep[meteoritic value, e.g.][]{balsiger1968}, to reflect the composition of the interstellar medium. However, $^6$Li is destroyed at temperatures greater than 2 million K, while $^7$Li is destroyed at 2.5 million K, and is therefore the more fragile of the two, making it even more prone to stellar astration. In practise, the $^6$Li contribution to the 670.7\,nm feature is typically very small for non-active stars \citep[e.g.][]{asplund2006,lind2013}, and at the resolution and $S/N$ typical for spectra of the current sample it will be nearly impossible to detect the lighter isotope, if at all present.

\begin{figure}
\resizebox{\hsize}{!}{
        \includegraphics[viewport=0 0 504 355,clip]{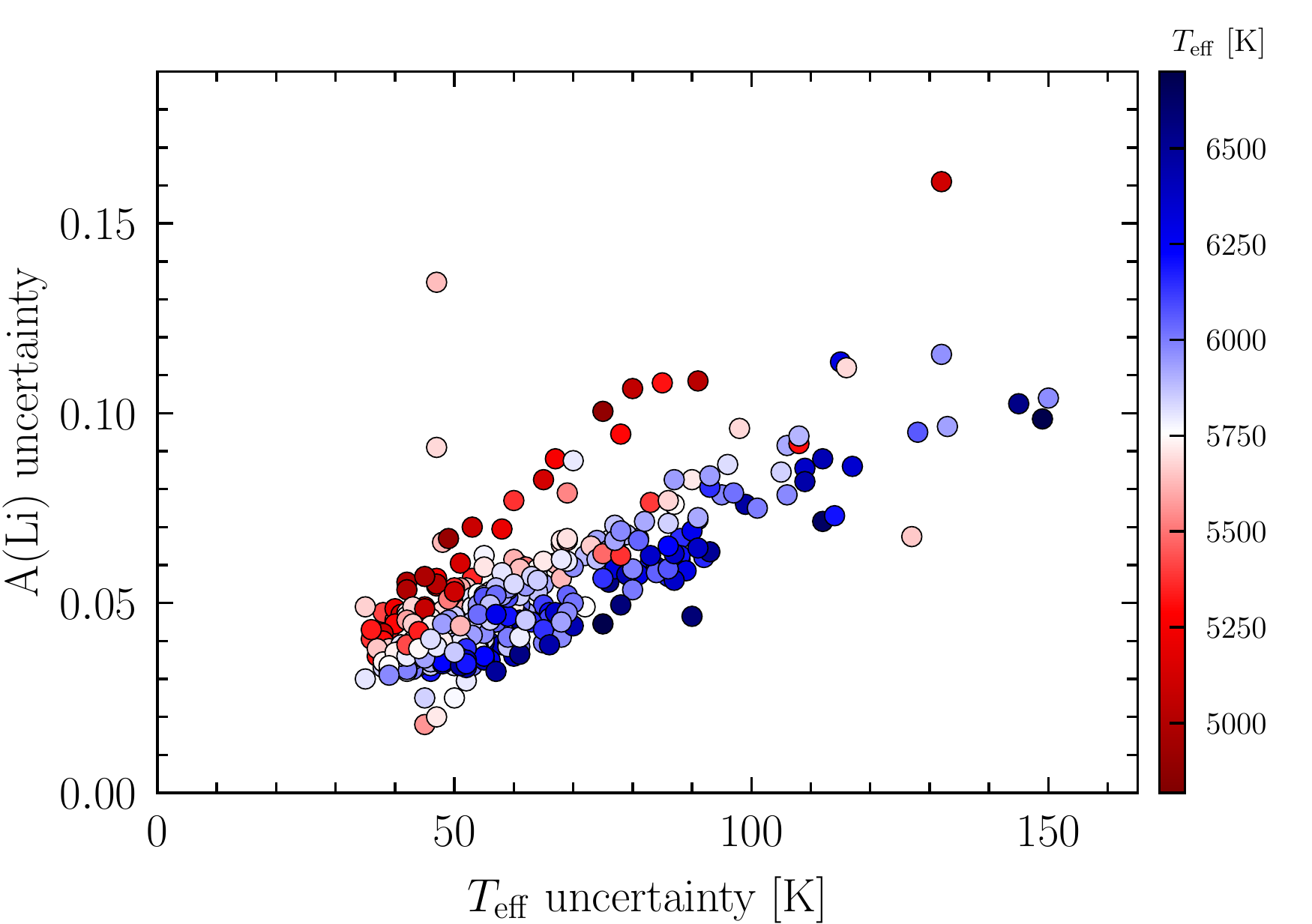}}
\caption{Uncertainties in the derived Li abundances due to the uncertainties in the effective temperatures. The stars have been colour-coded according to their effective temperatures.
}
\label{fig:uncertainties}
\end{figure}

There is a blending Fe\,{\sc i} line at 670.743\,nm that is irrelevant at low metallicity, but that has to be included when analysing metal-rich dwarf stars. The list of atomic lines used in the synthesis is given in Table~\ref{tab:linelist}. The $\log gf$ values for the different Li components are taken from \cite{smith1998} and the wavelength and $\log gf$-value for the blending Fe\,{\sc i} line from \cite{nave1994} and for the other lines from \cite{melendez2012}. Atomic data for other weak spectral lines farther away the $^7$Li line were queried from the VALD database \citep{vald_1,vald_2,vald_3,vald_4}. 

Synthetic spectra were calculated with version 298 of the SME software \citep[original version 1.0 described in][]{valenti1996} together with the MARCS model stellar atmospheres \citep{gustafsson2008}. Note that SME was simply used as a spectrum synthesiser without the fitting routines that can be used to find stellar parameters and abundances. We used our own fitting routines, and the choice of using SME as a synthesiser is because it contains a grid of the model atmospheres and an interpolator allowing easy access to model atmospheres with any combination of stellar parameters.

In addition to atomic line broadening, the observed line profile is broadened by the instrument, the line-of-sight component of the stellar rotation ($v_{\rm rot} \sin i$), and large-scale motions in the stellar atmosphere (macroturbulence, $v_{\rm macro}$). The instrument broadening is set by the resolving power of the spectrograph ($R$) and is treated with a Gaussian profile, while $v_{\rm rot} \sin i$ and $v_{\rm macro}$ are jointly accounted for with a radial-tangential (RAD-TAN) profile. To determine the RAD-TAN broadening we used the Fe\,{\sc i} lines located at 606.5, 654.6, and 667.8\,nm. These lines are unblended, far from saturated, and roughly on the same part of the curve-of-growth as the lines of interest for our synthesis. The Li abundance was then determined through a simple $\chi^2$ minimisation routine, that is illustrated in Fig.~\ref{fig:hip1931}. Lastly, NLTE corrections from \cite{lind2009} were added to the Li abundances.

In total we were able to clearly detect the Li line at 670.7\,nm and determine Li abundances for 420 stars. For another 121 stars we could estimate upper limits to the Li abundance. For the remaining 173 stars the spectra were not of sufficient quality or contained artefacts, and no Li abundances are reported for those.

\begin{figure}
\resizebox{\hsize}{!}{
        \includegraphics[viewport=0 0 514 290,clip]{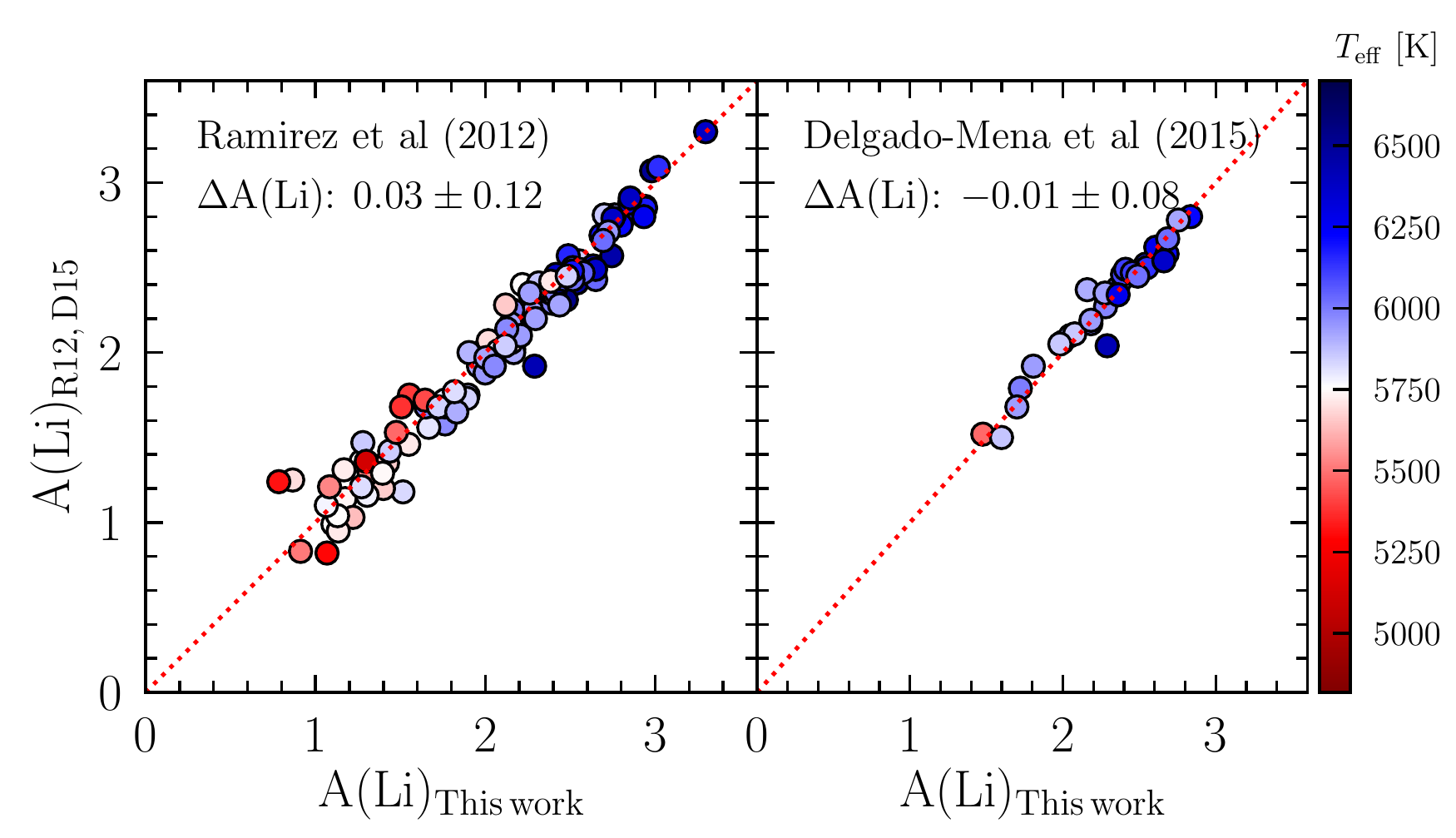}}
\resizebox{\hsize}{!}{
        \includegraphics[viewport=0 0 514 290,clip]{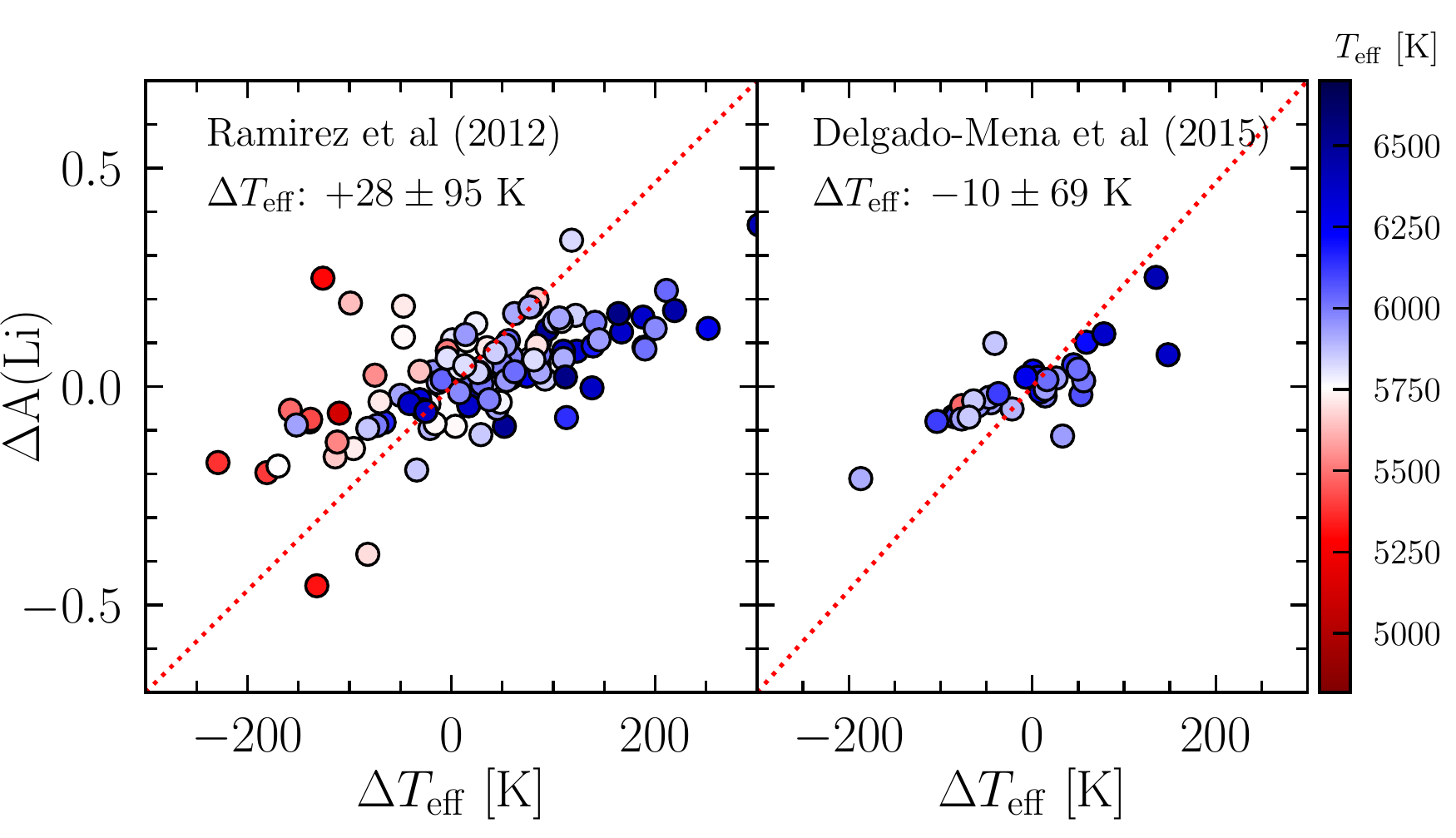}}
 \caption{The plots on the left-hand side show comparisons of Li abundances and effective temperatures for 117 stars in common with \cite{ramirez2012} and the plots on the right-hand side comparisons for 33 stars in common with \cite{delgadomena2015}. The red dotted lines shows the one-to-one relationships, and the average differences ($\Delta$A(Li) and $\Delta\teff$, calculated as our values minus theirs) and one sigma dispersions are indicated in the plots. The stars have been colour-coded according to their effective temperatures.
 }
\label{fig:uncertainties2}
\end{figure}

\begin{table*}
\setlength{\tabcolsep}{1.8mm}
\centering
\caption{Li abundances for the 714 stars\tablefootmark{$\dagger$}
}
\footnotesize
\label{tab:parameters}
\begin{tabular}{cccccccccccccccc}
\hline\hline
\noalign{\smallskip}
HIP   &
$\teff$ &
$\epsilon\teff$ &
$\log g$ &
$\epsilon\log g$ &
M &
M$^{\rm high}$ &
M$^{\rm low}$ &
Age &
Age$^{\rm high}$ &
Age$^{\rm low}$ &
A(Li)$_{\rm NLTE}$ &
A(Li)$_{\rm NLTE}^{\rm high}$ &
A(Li)$_{\rm NLTE}^{\rm low}$ &
$\Delta_{\rm NLTE}$ &
Flag \\
\noalign{\smallskip}
\hline
\noalign{\smallskip}
  1931 & & & & & & & & & & && & & & \\
\noalign{\smallskip}
\hline
\end{tabular}
\tablefoot{
\tablefoottext{$\dagger$}{
For each star we give effective temperature and surface gravity and their estimated uncertainties. For the masses and ages we give the best value and the give the lower and upper estimates. These values are taken from \cite{bensby2014}. Then we give the NLTE corrected A(Li) abundance and the low and high values determined by changed the effective temperatures by their uncertainties. The Li NLTE corrections that were added are also given. The last column is a flag where the value "0" means that it is a value from a well-fitted line. A value of "1" means that the Li abundance is an upper limit.
The table is only available in electronic form at the CDS via anonymous ftp to
\url{cdsarc.u-strasbg.fr (130.79.128.5)} or via
\url{http://cdsweb.u-strasbg.fr/cgi-bin/qcat?J/A+A/XXX/AXX}
}}
\end{table*}

\subsection{Li uncertainties}

Li abundances are very sensitive to changes in $\teff$, while hardly at all to changes in $\log g$ and [Fe/H]. Hence, the uncertainty in $\teff$ is the dominant source to the uncertainties in the derived Li abundances.
We have therefore redone the synthesis of the Li feature for all stars with the errors in the effective temperatures, from \cite{bensby2014}, applied, giving a lower and an upper uncertainty on each Li abundance based on the temperature uncertainty.  Figure~\ref{fig:uncertainties2} shows how the A(Li) uncertainty varies with the uncertainty in effective temperature. For the great majority of the stars in the sample the A(Li) uncertainties are well below 0.1\,dex. It should be noted that in the upcoming plots where error bars are included these uncertainties are usually smaller than the sizes of the markers due to the wide range of Li abundances that the sample spans.

Table~\ref{tab:parameters} gives the stellar parameters, NLTE corrected Li abundances, the upper and lower Li abundances (based on uncertainties in $\teff$), and the NLTE corrections that were added to the LTE abundances.

\subsection{Comparison to other studies}

Figure~\ref{fig:uncertainties2} shows comparisons between our Li abundances and temperatures to the Li abundances and temperatures from two other studies: \cite{ramirez2012} that have 117 stars in common with our sample, and \cite{delgadomena2015} that have 33 stars in common with our sample. The agreements are good, on average our Li abundances are 0.03\,dex higher than the \cite{ramirez2012} values, with a one-sigma dispersion of 0.12\,dex. The slight offset is likely due to slight differences in the effective temperatures, as can be seen in the bottom left plot of Fig.~\ref{fig:uncertainties2}. On average our effective temperatures are 28\,K higher, with a one-sigma dispersion of 95\,K. There is, however, a tendency that for cooler stars, our temperatures and Li abundances are lower, while for hotter stars our temperatures and Li abundances are higher. This could possibly be traced to the methods by which the temperatures were inferred, we used excitation balance, while \cite{ramirez2012} based their temperatures on the infrared flux method. The A(Li) differences are however not large enough to delve deeper into this.

The right-hand side of Fig.~\ref{fig:uncertainties2} shows a similar comparison to 33 stars that we have in common with \cite{delgadomena2015}. Again, the agreement is very good in both derived Li abundances as well as effective temperatures, where our temperatures are on average 10\,K lower with a one-sigma dispersion of 69\,K. Also here we see a trend of $\teff$ being the main driver of the differences in Li, that most likely is due to differences in the methods used to determine stellar parameters. Again the differences are too small to investigate this further.

There are many more studies of Li in the literature \citep[e.g.][]{lopezvaldivia2015,gonzalez2015,pavlenko2018}. One should however be aware of that studies of Li usually use absolute abundances, that is, not relative to (or normalised to)  the abundance of a standard star such as the Sun. Therefore the linelists and atomic data of blending lines are important.  We experimented in varying the atomic data (the $\log gf$-values) for some of the blending lines, and in particular for metal-rich stars where lines become stronger, and found none or minuscule changes on our Li abundances (third decimal on the Li abundance). Hence, as we do not see any large differences, and as we in general reproduce the depth of the blending \ion{Fe}{i} line located just to the left of the Li feature, and as our Li results are perfectly in line with the two studies illustrated in Fig.~\ref{fig:uncertainties2}, we deem it unnecessary to present a comparison to all. The differences that are present are very small given the wide range of Li abundances that are observed.

\begin{figure}
\resizebox{\hsize}{!}{
        \includegraphics{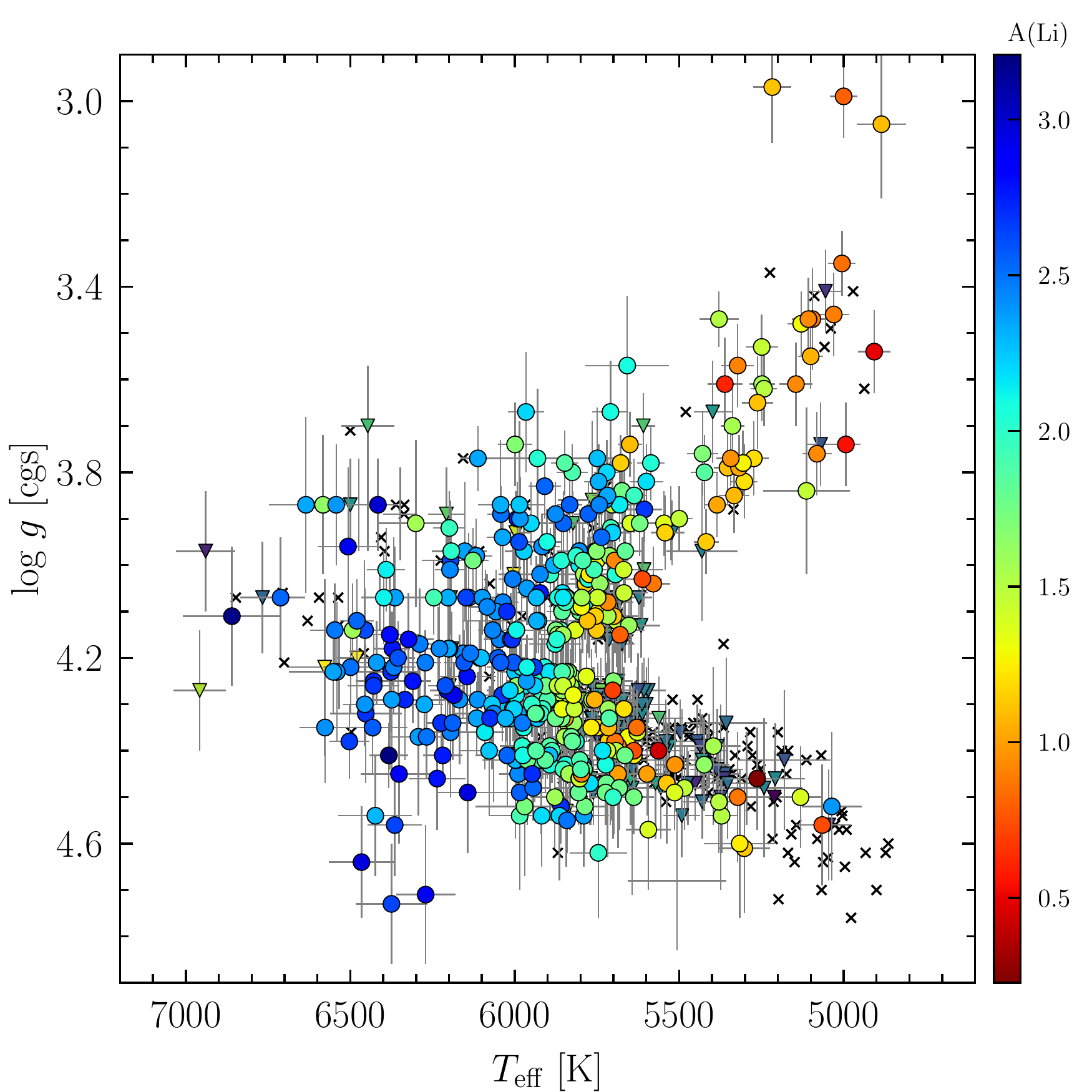}}
\caption{HR diagram  for the sample of 714 stars. Li abundances were determined for 420 stars (colour-coded circles), upper limits are reported for 121 stars (triangles), and for 173 stars no Li abundances are reported at all (crosses).
}
\label{fig:hr}
\end{figure}

\section{Li results}

An HR diagram for the sample is shown in Fig.~\ref{fig:hr} where the data points have been colour-coded based on their Li abundances. There is a clear Li gradient with temperature, with the lowest Li abundances on the lower main sequence and on the red giant branch. The highest Li abundances are found on the upper main sequence and around the turn-off. There is a slight gap in the HR diagram around $\log g \approx 4.1-4.2$ that is artificial due to an empirical correction that was applied to the stellar parameters because of an un-resolved issue with the stellar parameters based on ionisation balance \citep[see Fig.~11 in][]{bensby2014}. These corrections had some effects on $\log g$ and only minor to $\teff$. As Li abundances are not sensitive to changes in $\log g$ these corrections are not important for the Li results of the current study.

\begin{figure*}
\resizebox{\hsize}{!}{
        \includegraphics[viewport=0 0 504 350,clip]{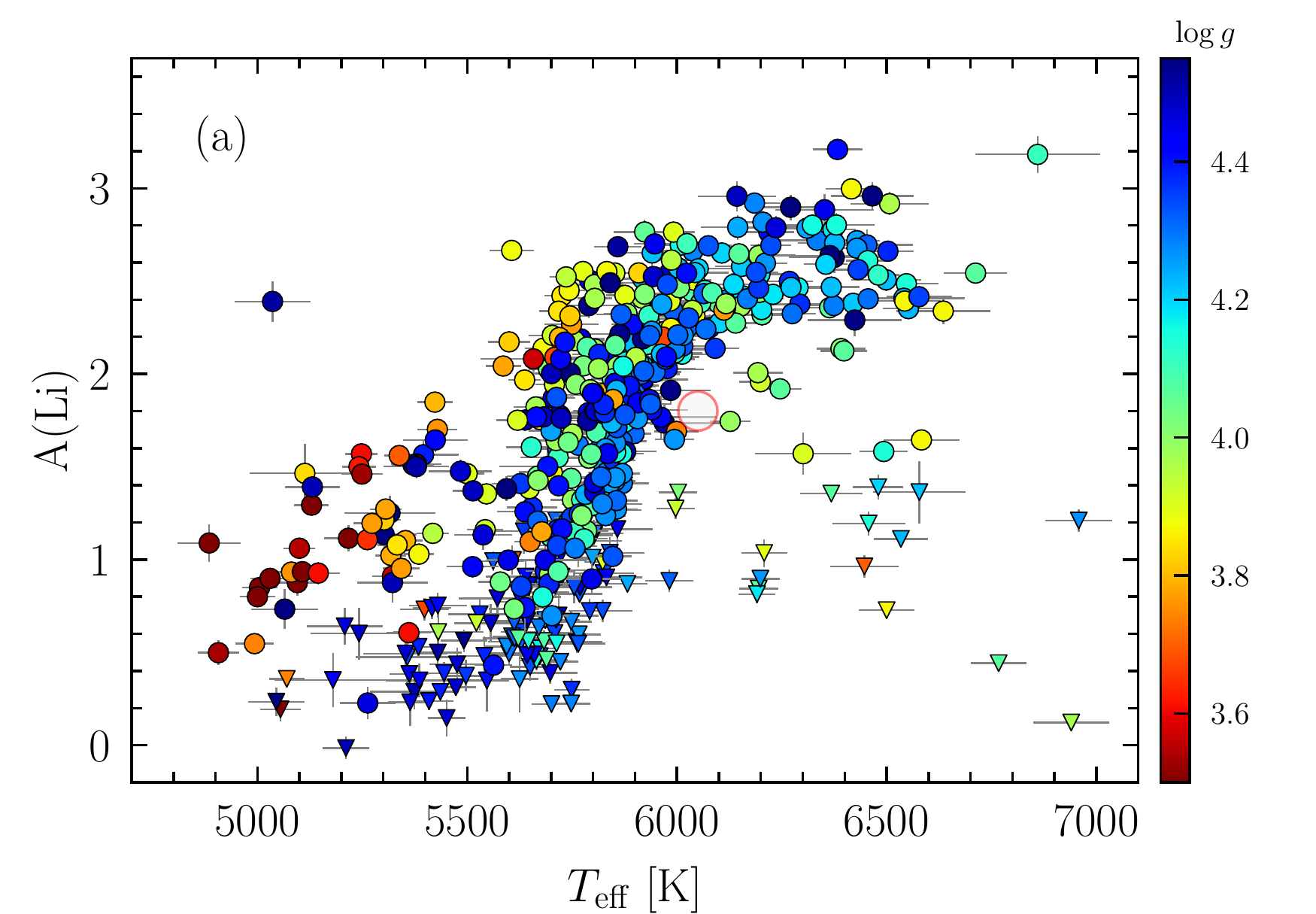}
        \includegraphics[viewport=0 0 504 350,clip]{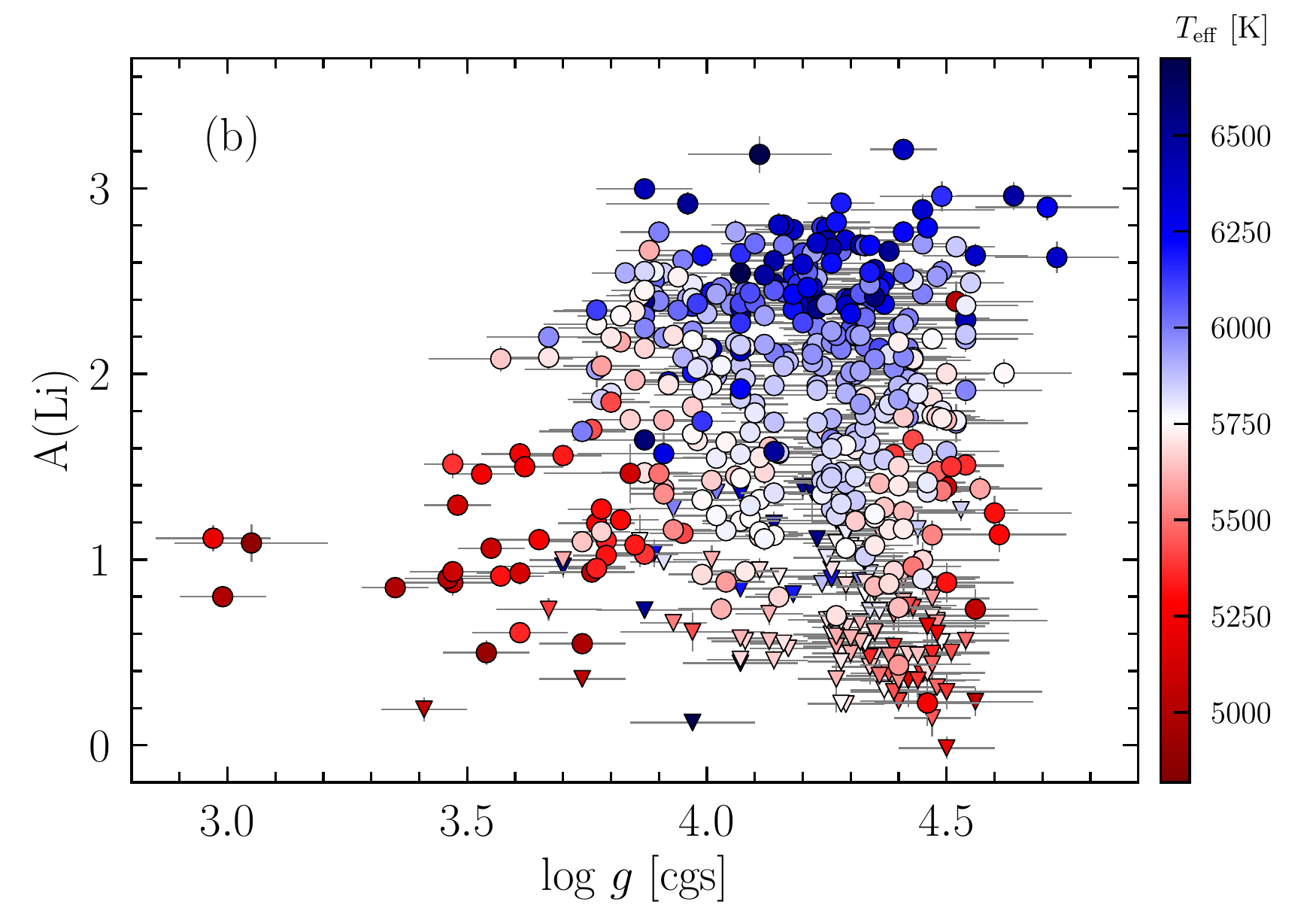}}
\resizebox{\hsize}{!}{
        \includegraphics[viewport=0 0 504 350,clip]{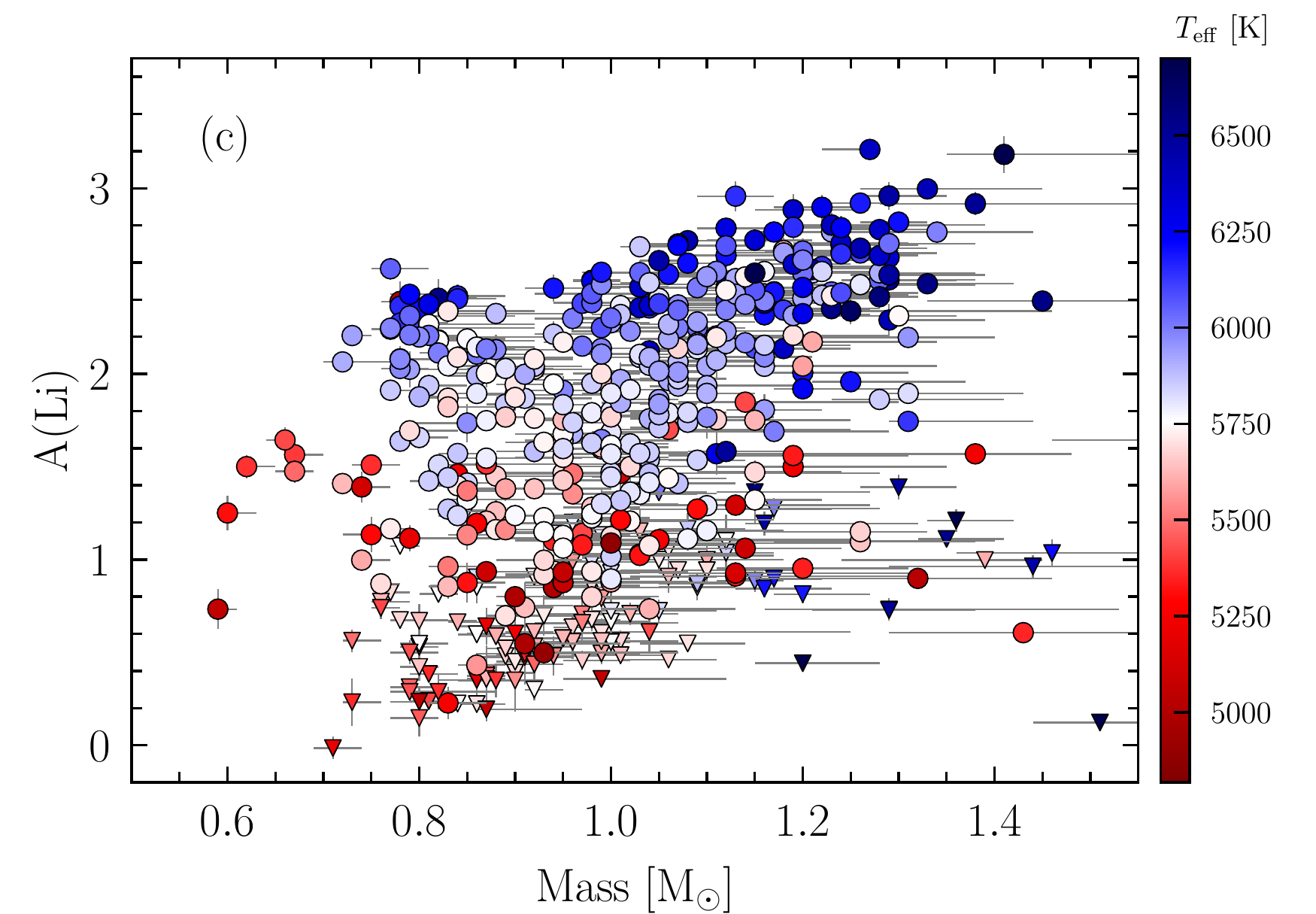}
        \includegraphics[viewport=0 0 504 350,clip]{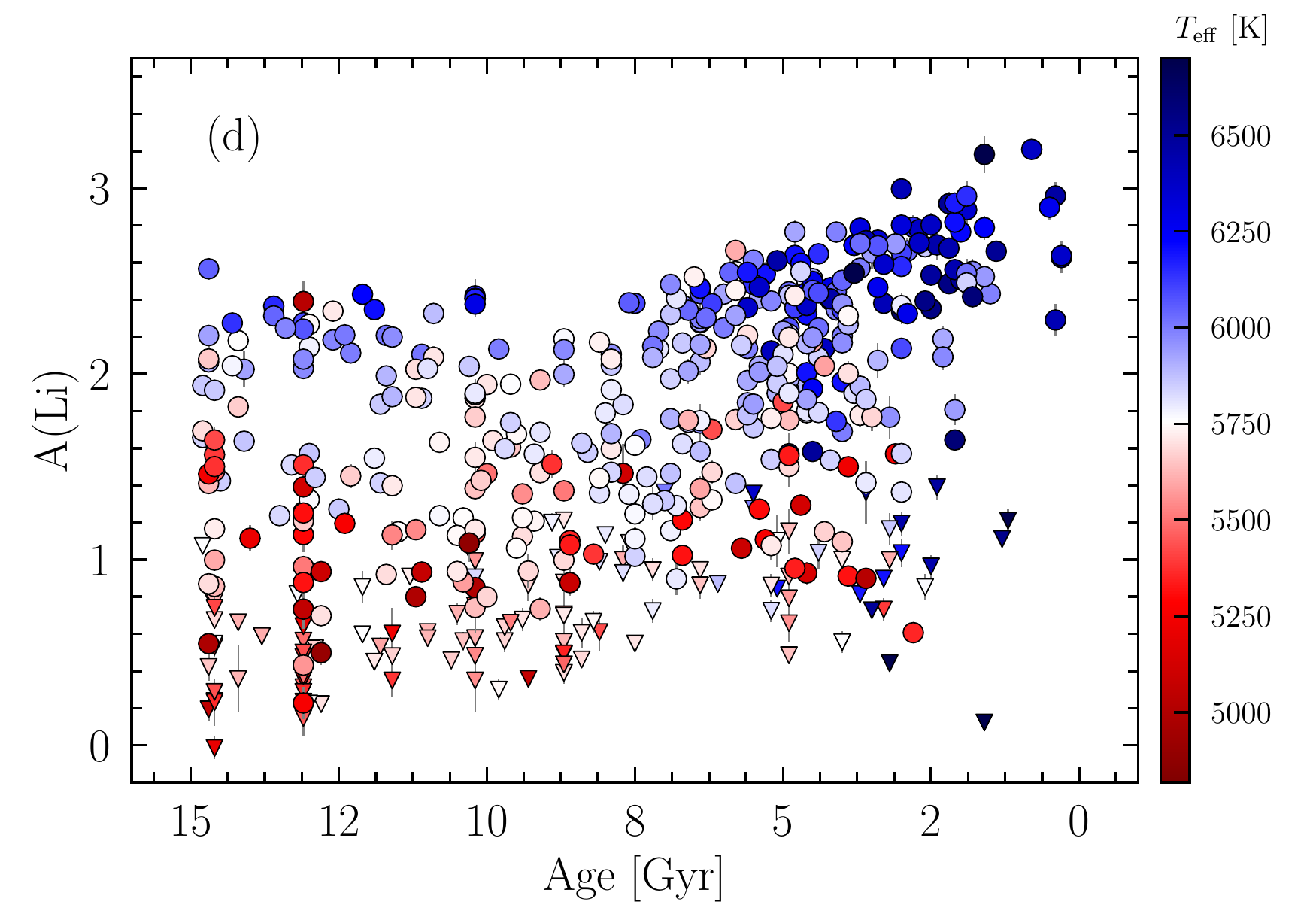}}
\resizebox{\hsize}{!}{
        \includegraphics[viewport=0 0 504 350,clip]{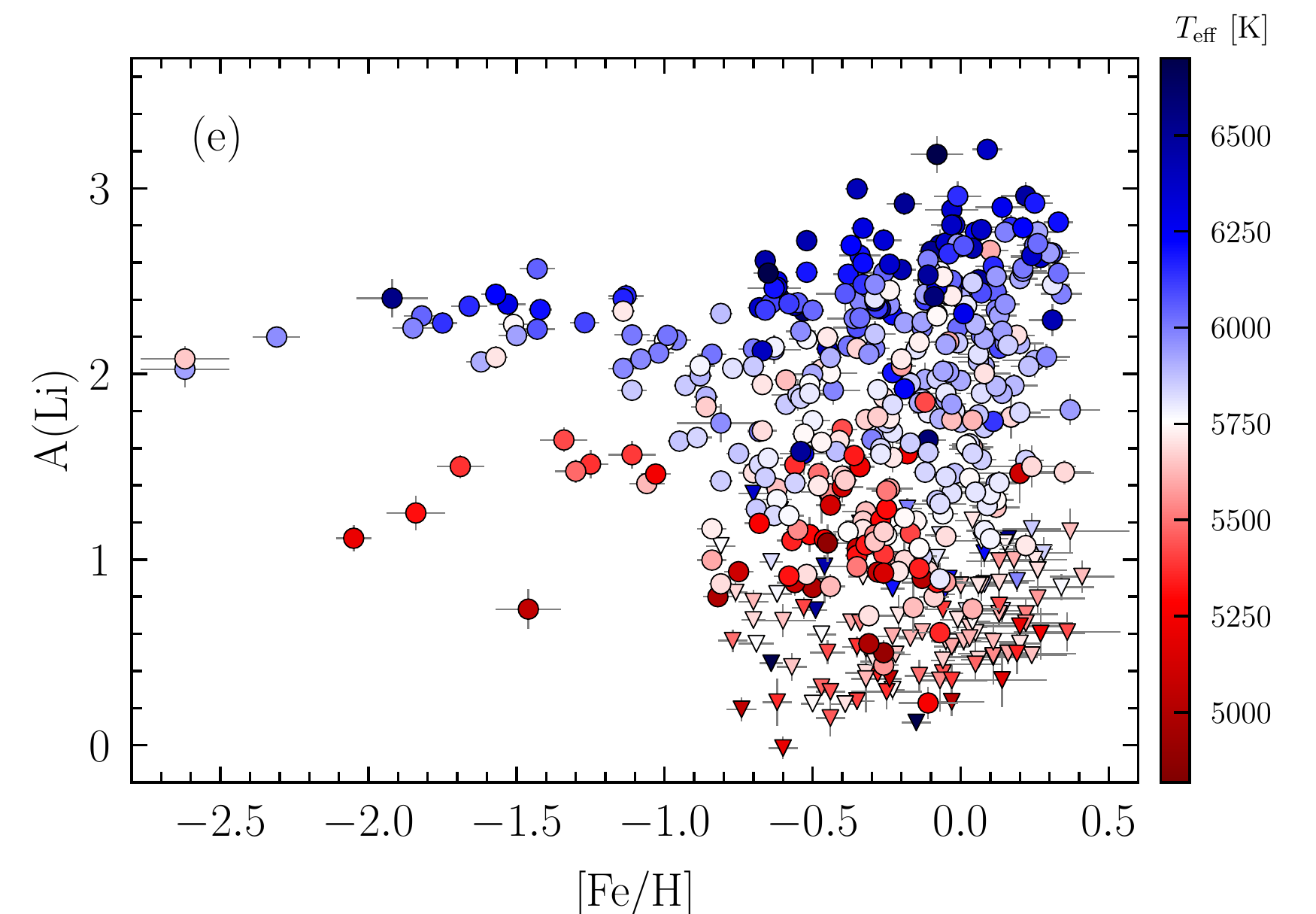}
        \includegraphics[viewport=0 0 504 350,clip]{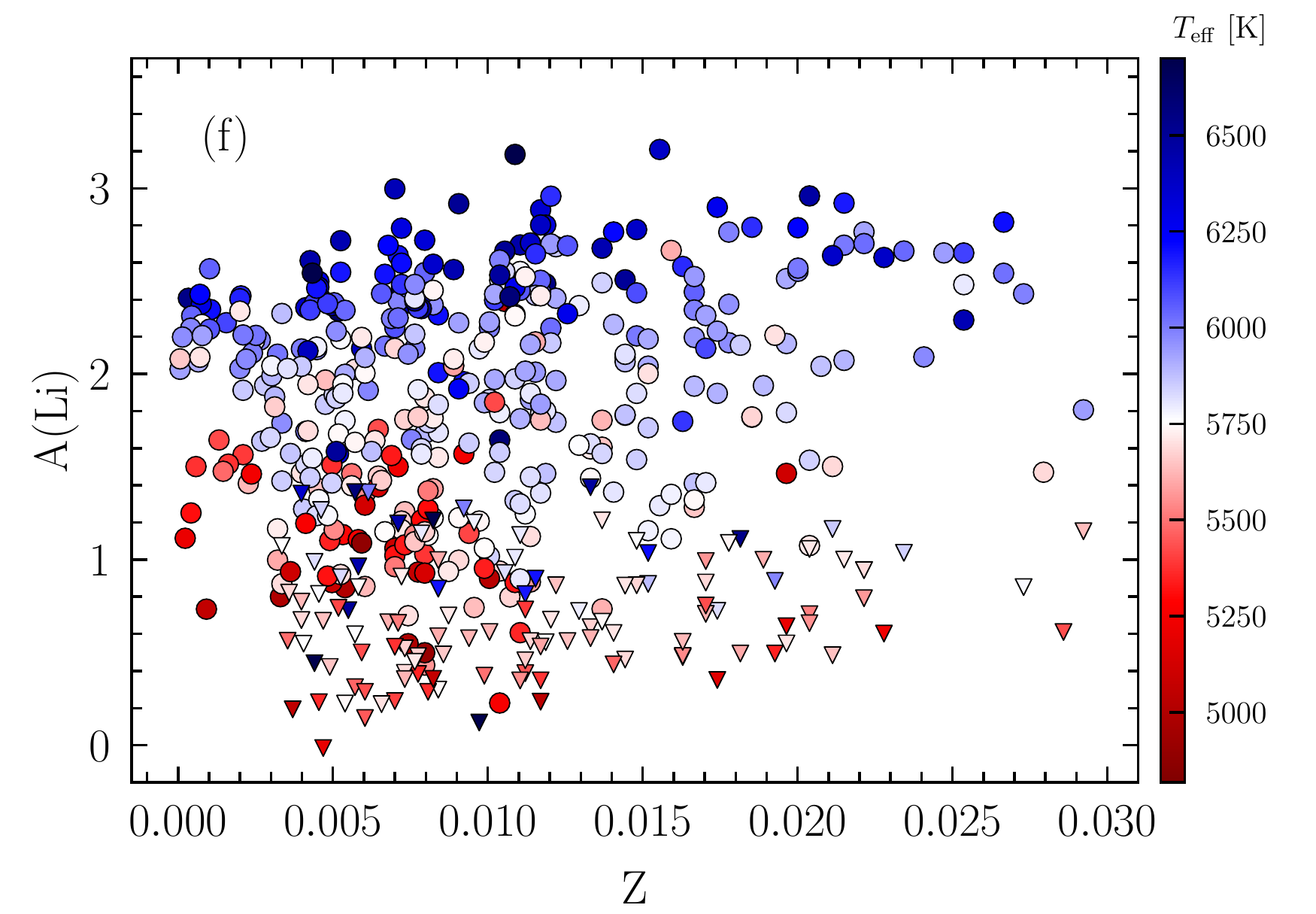}}
\caption{Li abundances versus effective temperature and metallicity. Black circles represent stars classified as main sequence stars and the open blue circles stars that passed the turn-off point (in total 420 stars, see Fig.~\ref{fig:hr}). The plots also include the 121 stars that only have upper limit Li abundances (grey triangles).}
\label{fig:litrends}
\end{figure*}

Figure~\ref{fig:litrends} shows how the Li abundances of the full sample vary with stellar effective temperature, surface gravity, metallicity, mass, and age. As expected in such a broad parameter range, Li spans over several orders of magnitude in abundance and display complex correlations on the dependent variables.

\subsection{Li versus effective temperature}

The evolution of A(Li) with $\teff$ shows the typical appearance, with the A(Li) trend  decreasing with decreasing temperatures. This is due to the larger convection zones in cooler stars allowing for a higher degree of Li depletion. At temperatures above about 5900\,K the A(Li) trend levels out on a temperature-independent band with full range between $\rm A(Li)\approx2.3-3.0$, that is, between the typical Spite plateau value and close to the meteoritic value. There are also a dozen or so outliers from the main trend, with atypically low Li abundances. There is no clear reason to these, they are not rapid rotators, or show any signatures of binarity in their spectra. 

The stars at the lower temperature end have mainly upper limit estimates of the Li abundances. This is due to that the Li line becomes weaker and the SNR in the spectra does not allow us to determine the Li abundance with good accuracy. In this range there are some stars with lower surface gravities that have higher Li abundances. These stars are subgiants and have higher masses than the main-sequence stars of same temperature, meaning they preserved more Li on the main sequence and now start to experience the post-MS depletion.

\cite{ramirez2012} reported a 'Li desert' around $\teff\approx6050$\,K and $\rm A(Li)\approx1.8$ (marked by the grey circle with red edges in Fig.~\ref{fig:litrends}). Surprisingly, we cannot find any stars in this small region either. Recently, \cite{aguileragomez2018} investigated this Li desert in detail and concluded that the stars below the desert have evolved from the Li dip, and hence have lower Li abundances compared to other stars at this temperature, above the desert, that have higher Li abundances.

\subsection{Li versus surface gravity}

The A(Li) abundance pattern with $\log g$ shows no clear pattern in the range $3.8\lesssim\log g\lesssim4.6$. The hottest stars show the highest Li abundances and the coolest stars the lowest. Again a manifestation of the destruction of Li in the deeper convective envelopes of the cooler stars. For $\log g\lesssim3.8$ we note that we have no stars with as high Li abundances as for the higher surface gravities. This is because the sample does not contain hotter main sequence stars at lower surface gravities (i.e. no massive young stars).

\subsection{Li versus mass and age}

The correlation of  A(Li) with stellar mass is shown in Fig.~\ref{fig:litrends}c. Low-mass stars have a lower maximum Li abundance compared to the higher-mass ones. Note, however, that there is also a strong correlation between stellar mass and age, the high-mass stars being the younger stars. This indicates that the Li abundance increases with decreasing age, which is also supported by Fig.~\ref{fig:litrends}d where A(Li) is plotted versus stellar age. It appears as if the maximum Li abundance stays essentially flat the first few billion years and around eight billion years ago A(Li) starts to increase. This appearance will be further discussed in Sect.~\ref{sec:thinthick} when discussing the evolution of Li in the Galactic thin and thick disks separately.

\begin{figure}
\centering
\resizebox{\hsize}{!}{
        \includegraphics{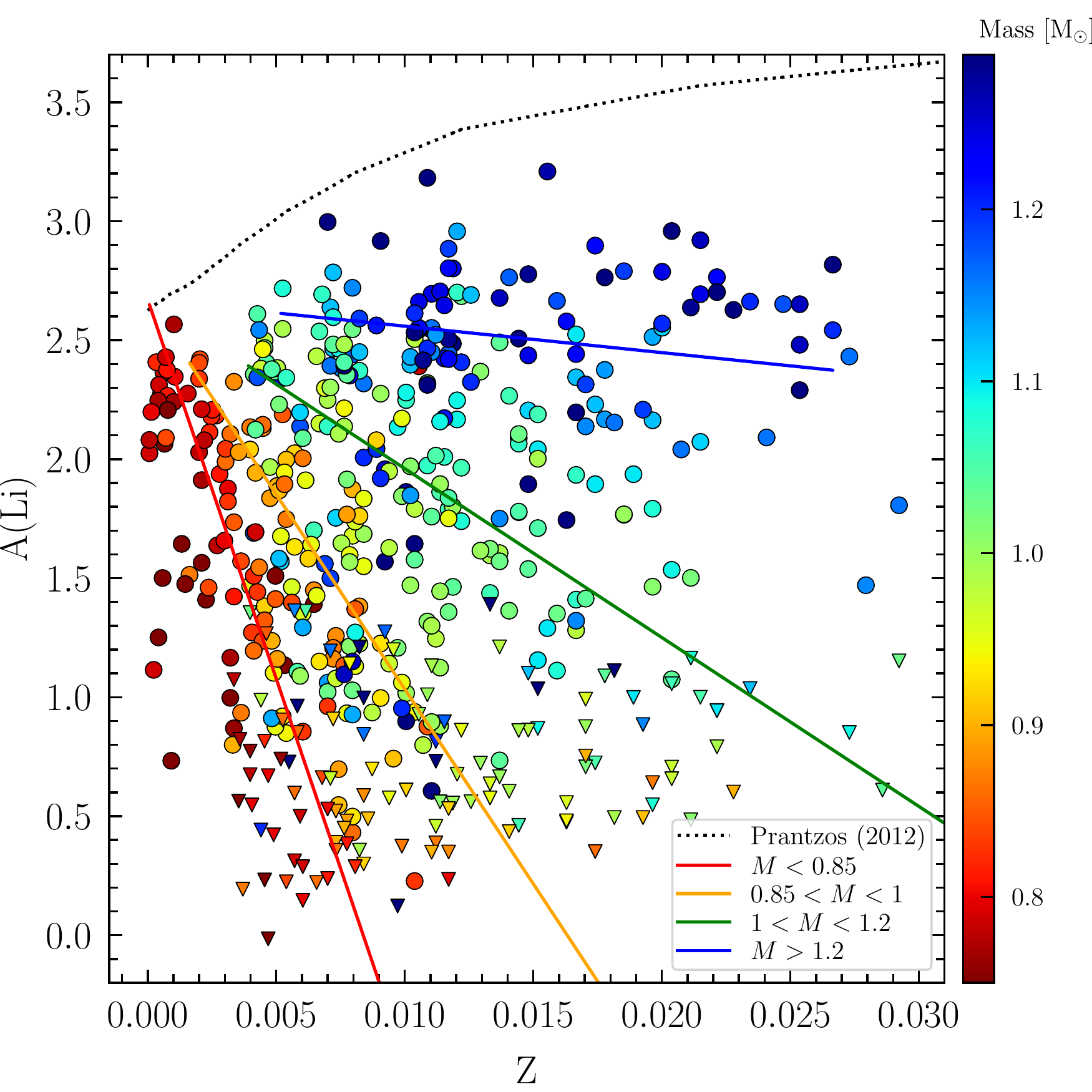}}
\caption{Li abundance versus stellar metallicity in mass bins for the 397 pre-turn-off stars cooler than 6500\,K.  The dashed lines at the top represent an approximate representation of the Galactic evolution of Li from \cite{prantzos2012}.
The metallicity $Z$ has been converted from [Fe/H] based on the relationship between [Fe/H] and $Z$ in the MARCS model stellar atmospheres.
}
\label{fig:lizmass}
\end{figure}

\subsection{Li versus metallicity}

The evolution of Li as a function of [Fe/H] is shown in Fig.~\ref{fig:litrends}e. At metallicities below $\rm [Fe/H]\approx-1$ it levels out at a value of $\rm A(Li)\approx2.3$ which is consistent with the Li plateau that has been observed by many studies \citep{spite1982}. At higher metallicities there is a wide range of Li abundances demonstrating that Li has been both produced and destroyed in stars. If considering only warmer stars that are least subject to Li depletion we see that the A(Li) trend increases with metallicity, reaching values higher than three around solar metallicities.

In the most metal-rich stars at $\rm [Fe/H]\approx+0.1$, A(Li) appears to be levelling out, or might even be slightly decreasing. This was also seen in the studies by \cite{delgadomena2015}, \cite{guiglion2016}, and \cite{fu2018}. Such a decrease, if confirmed, is the first example of an element produced by stars that decreases during the last few billion years of the Galactic chemical evolution \citep{prantzos2017}. The trend is more pronounced on the linear metallicity scale shown in Fig.~\ref{fig:litrends}f, where by $Z=Z_\odot \times 10^{\rm [Fe/H]}$ and $Z_\odot=0.014$. The A(Li) trend increases from $Z=0$ to $Z\approx0.01$, after which it appears to level out, and possibly also to decrease.

\subsection{Li - metallicity - mass correlations}

Inspired by similar plots in \cite{nissen2012}, Fig.~\ref{fig:lizmass} shows how the Li abundances vary with metallicity for stars in different mass intervals. Our sample reveals a remarkably clear correlation between Li abundance, stellar mass, and (linear) metallicity.  Evidently, Li is anti-correlated with metallicity within each mass bin, with the steepest dependence for stars with the lowest masses. Note that in the absence of astration, Galactic production of Li would result in slopes of opposite sign, dictating an increase of the Li abundance with metallicity.  Assuming a primordial contribution of $A\rm(Li)=2.67\pm0.06$ \citep{cyburt2016}, the different sources of Li must have contributed with almost 0.6\,dex of Galactic Li to meet the meteoritic value $A\rm(Li)=3.26$  at solar metallicity \citep{lodders2009}. We illustrate the expected behaviour with an approximate Li evolution model from \cite{prantzos2012} that assumes the Galactic contribution to be linearly dependent on metallicity from $\rm[Fe/H]\approx-1.0$. Hence, while the highest mass stars show an approximately flat slope, this should be interpreted as the cancellation of two independent, competing dependencies; stellar depletion, which increases with increasing metal content for a given stellar mass, and Galactic production.

The linear regressions to the Li-metallicity relationship performed in each mass bin were forced to converge at $A\rm(Li)=2.67$ at zero metallicity, to emphasise how well compatible the sample is with a unique primordial abundance. While the origin was here set to correspond exactly to the expected value, it is apparent from the plots that a significantly higher or lower origin would describe the sample less well in this setting, which is very encouraging.

\subsection{Li - age correlation for solar twin stars}

\begin{figure}
\centering
\resizebox{\hsize}{!}{
        \includegraphics[viewport=0 0 504 405,clip]{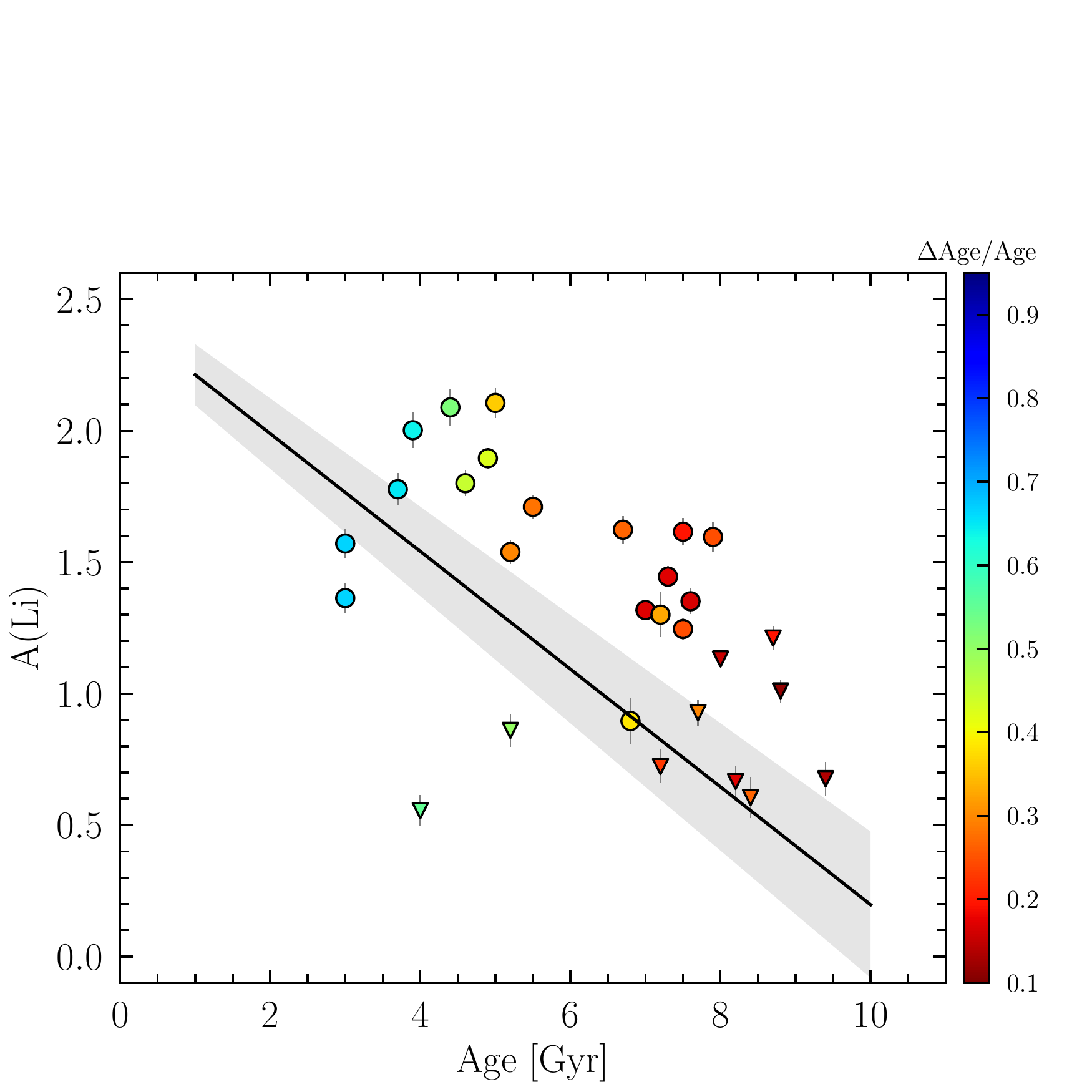}}
\caption{
Correlation between Li abundance and age for solar twin stars. The circles show stars that have well-determined Li abundances (19 stars) and the triangles those that only have upper limits on their Li abundances (10 stars). The definition of a solar twin is here: $\teff = 5690-5870$\,K, $\log g = 4.25-4.50$, and $\rm [Fe/H]=\pm0.11$. The colouring shows the relative age uncertainties of these stars (that can be substantial as these solar twin stars are located on the upper main sequence). The diagonal line shows the relationship between Li and stellar age from \cite{carlos2016}.
}
\label{fig:liagetwins}
\end{figure}

Recent studies have found a strong correlation between age and Li abundance for solar twin stars, meaning stars that have stellar parameters very similar to the Sun \citep{monroe2013,melendez2014,carlos2016}. They find that during the last 8 to 9 Gyr, the surface Li abundances of these stars have decreased by almost 2\,dex, which is interpreted as a signature of the the gradual destruction with time of Li due to that diffusion and mixing allows Li to be transported from the bottom of the convective zone to regions where it is destroyed. Figure~\ref{fig:liagetwins} shows the age-Li correlation for stars in our sample that have stellar parameters similar to the Sun. Here we used the same range of stellar parameters as \cite{carlos2016}, that is $\teff = 5690-5870$\,K, $\log g = 4.25-4.50$, and $\rm [Fe/H]=\pm0.11$. This left us with 19 stars with well-determined Li abundances and 10 stars with upper limit Li abundance estimations. As is seen in Fig.~\ref{fig:liagetwins} our results also reveal a potential Li-age correlation, although not as tight as the one seen in the studies listed above, and that clearly is offset to higher Li abundances relative to the relation found by \cite{carlos2016}. The average offset to the relation by \cite{carlos2016} is $+0.32$\,dex. The larger scatter is probably because of the larger age uncertainties for these stars that are located in a region of the HR diagram where the isochrones are not well separated. The other studies have smaller uncertainties on their stellar parameters as their spectra generally have signal-to-noise ratios higher than about 500, which results in better precision in stellar parameters and hence also stellar ages.  Why there is an offset of more than $+0.3$\,dex is unclear.
As stated by \cite{thevenin2017}, a scatter around the relation could be due to variations in the physical conditions during the pre-main sequence phase of the stars. That does not explain the offset though. We therefore deem the Li-age correlation for solar twin stars in our sample as dubious, and the relatively large age uncertainties for this subset of stars does not allow us to delve deeper into the causes.

\begin{figure}
\resizebox{\hsize}{!}{
        \includegraphics{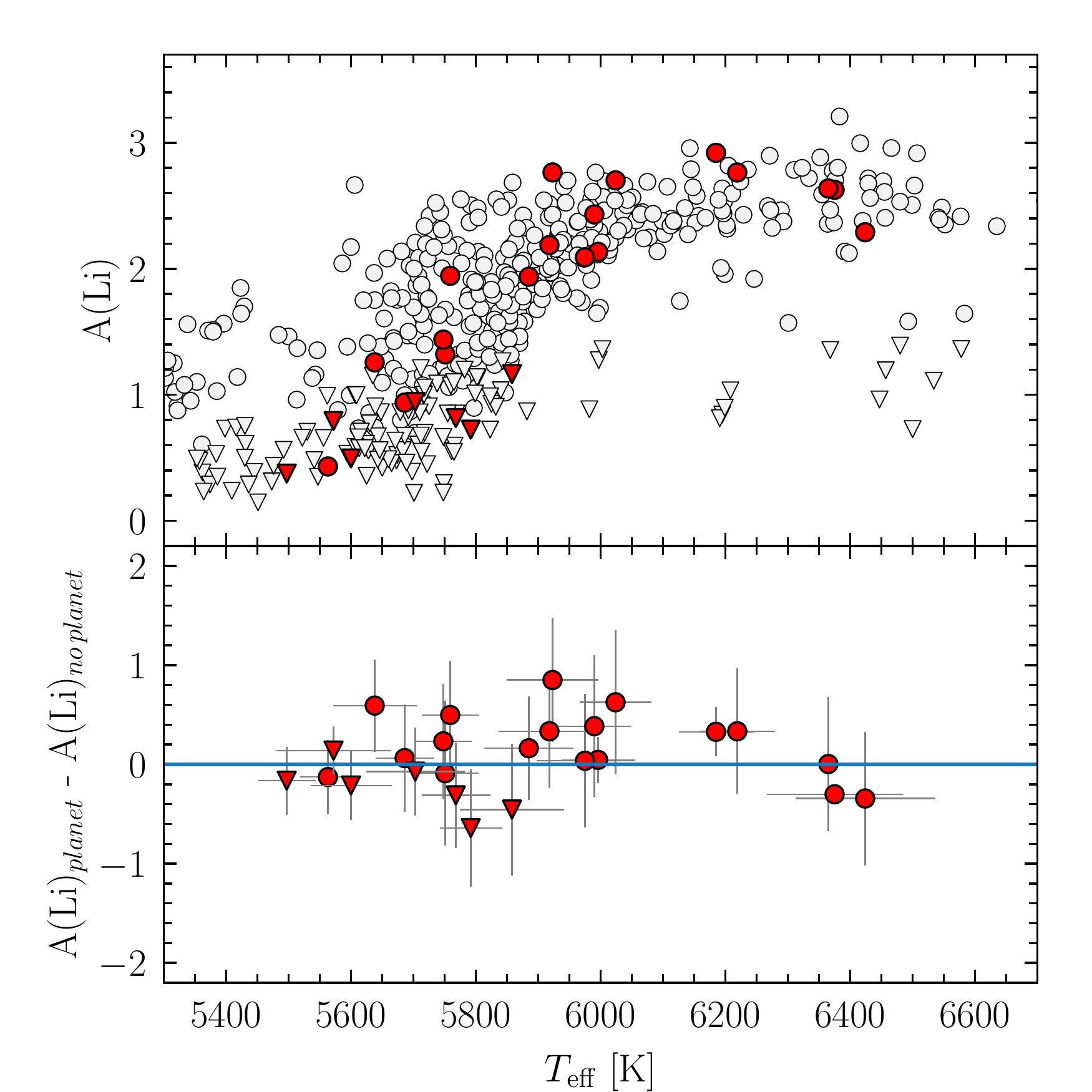}}
 \caption{The red circles in the upper panel marks the 18 stars in the sample that have exoplanets detected and good Li measurements. The bottom panel shows the difference between the A(Li) for individual stars to the mean A(Li) for stars that have no exoplanet detections and that have effective temperatures, surface gravities, and metallicities, within $\pm 75$\,K, $\pm 0.2$\,dex, and $\pm 0.2$\,dex. For stars with $\teff>6100$\,K the temperature range is expanded to $\pm200$\,K.
 }
\label{fig:planets}
\end{figure}

\subsection{Li in stars with planets}

Since \cite{king1997li} claimed a possibility that the low Li abundances in the Sun and 16~Cyg~B relative to 16~Cyg~A could be due to the presence of planetary companions in the former two, several studies have presented disparate results regarding if stars with detected planets have different Li abundance patterns compared to those with so far no detected planets. Studies that support the idea that there is no difference include \cite{ryan2000,luck2006b,baumann2010,ghezzi2010li,ramirez2012} and \cite{pavlenko2018}, while those that find that indeed there is a difference include \cite{israelian2004,takeda2005_li,chen2006,gonzalez2008,israelian2009,figueira2014,ggonzalez2014,ggonzalez2015,delgadomena2014} and \cite{mishenina2016}.

To see whether this study can add some further clues to this controversy the \url{http://exoplanets.org} website  \citep{han2014} was queried on February 14, 2018. We find that 36 stars in our sample have detected exoplanets, and that 18 of those have well-determined Li abundances and that 6 have upper limit Li detections. These $18+6$ stars are marked in red in Fig.~\ref{fig:planets}. They span a range of effective temperatures from about 5600\,K to 6500\,K. The bottom panel in Fig.~\ref{fig:planets} shows the differences between the Li abundance of the exoplanet stars and the mean Li abundance of other stars in the sample that have effective temperatures within $\pm75$\,K, $\log g$ within $\pm0.25$\,dex, and $\rm [Fe/H]$ within $\pm0.25$\,dex of each exoplanet star. Due to the low number statistics of stars with similar atmospheric parameters at higher $\teff$, the temperature range was expanded to $\pm200$\,K for stars with $\teff>6100$\,K. As stars at these temperatures have less Li depleted atmospheres, this expansion should not affect the results. The stars with detected exoplanets have on average 0.08\,dex higher Li abundances than those without planet detections, with a dispersion of 0.36\,dex. Using the median instead gives a Li-enhancement of 0.04\,dex. These numbers are well within the uncertainties and our sample reveals no statistically significant difference in Li abundances between stars with our without exoplanet detections.

\begin{figure}
\resizebox{\hsize}{!}{
        \includegraphics{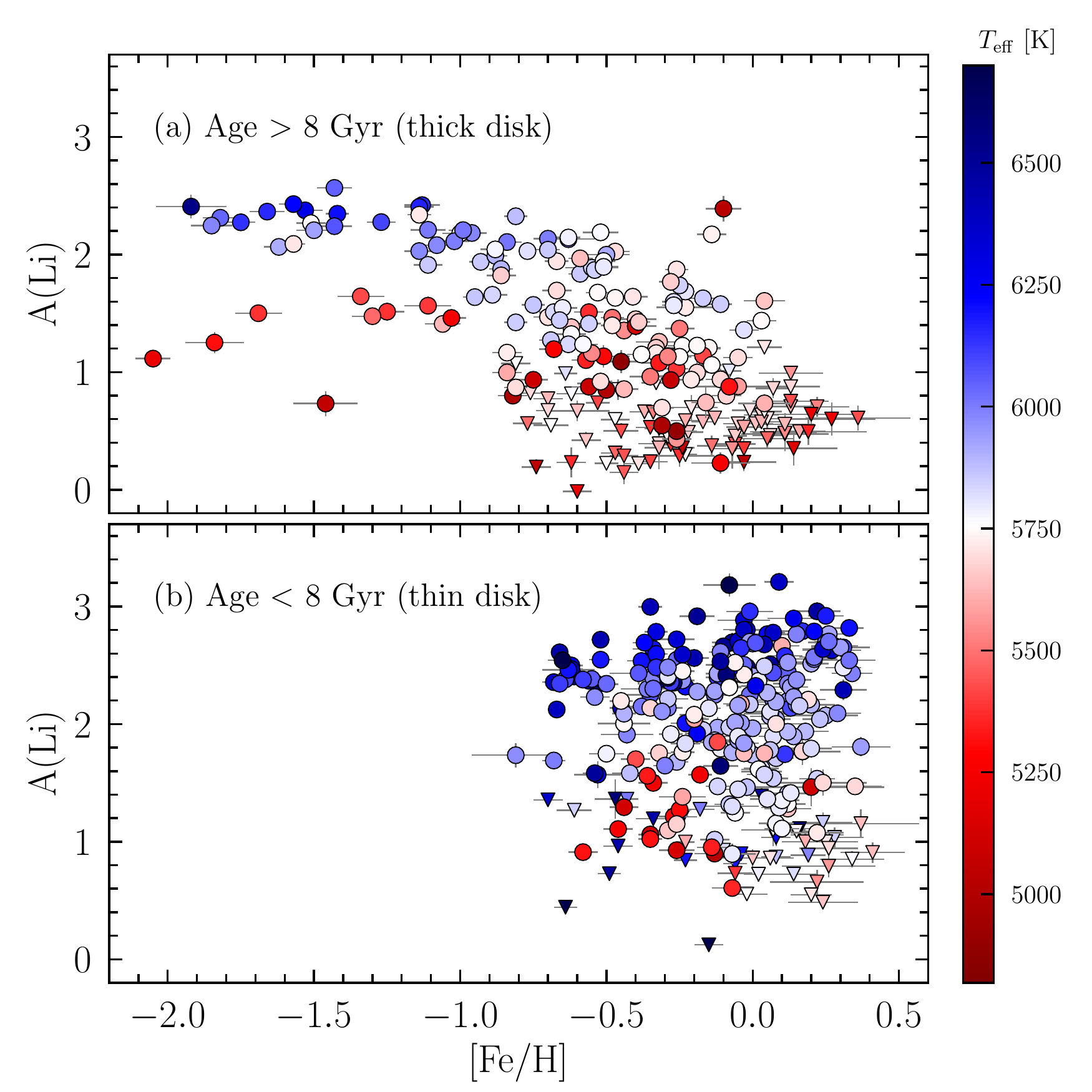}}
\caption{A(Li) versus [Fe/H] for the sample but divided into subsamples that are likely to be representative of the Galactic thin and thick disks. The separation has been done based on stellar ages and the thick disk sample (upper panel) contains stars older than 8\,Gyr, while the thin disk sample (lower panel) contains stars younger than 8\,Gyr. The colour-coding is based on the effective temperatures, as shown in the colour-bar on the right-hand side.
 }
\label{fig:thickdisk}
\end{figure}
\begin{figure*}
\resizebox{\hsize}{!}{
      \includegraphics[viewport=0 0 308 504,clip]{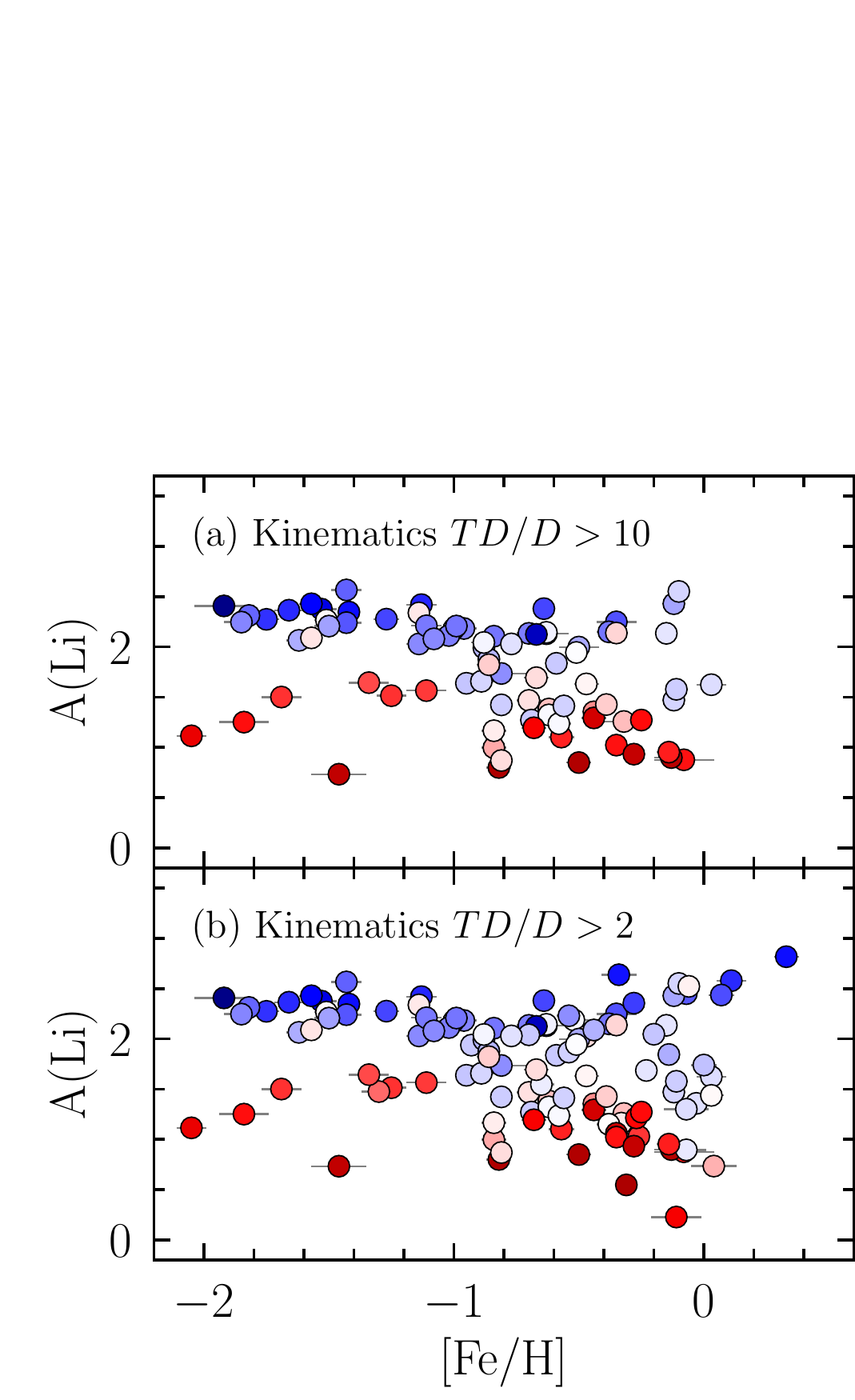}
      \includegraphics[viewport=0 0 308 504,clip]{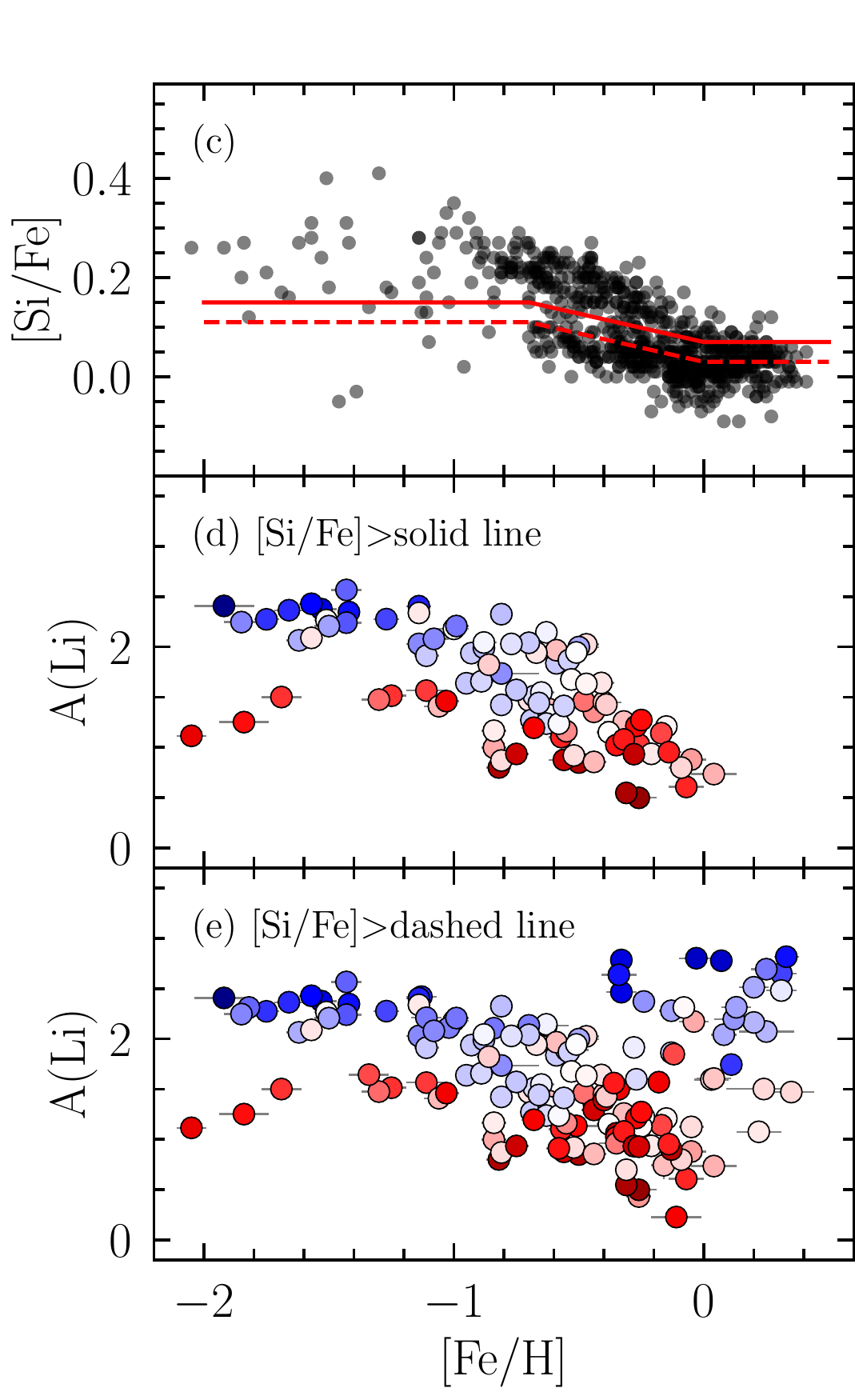}
      \includegraphics[viewport=0 0 370 504,clip]{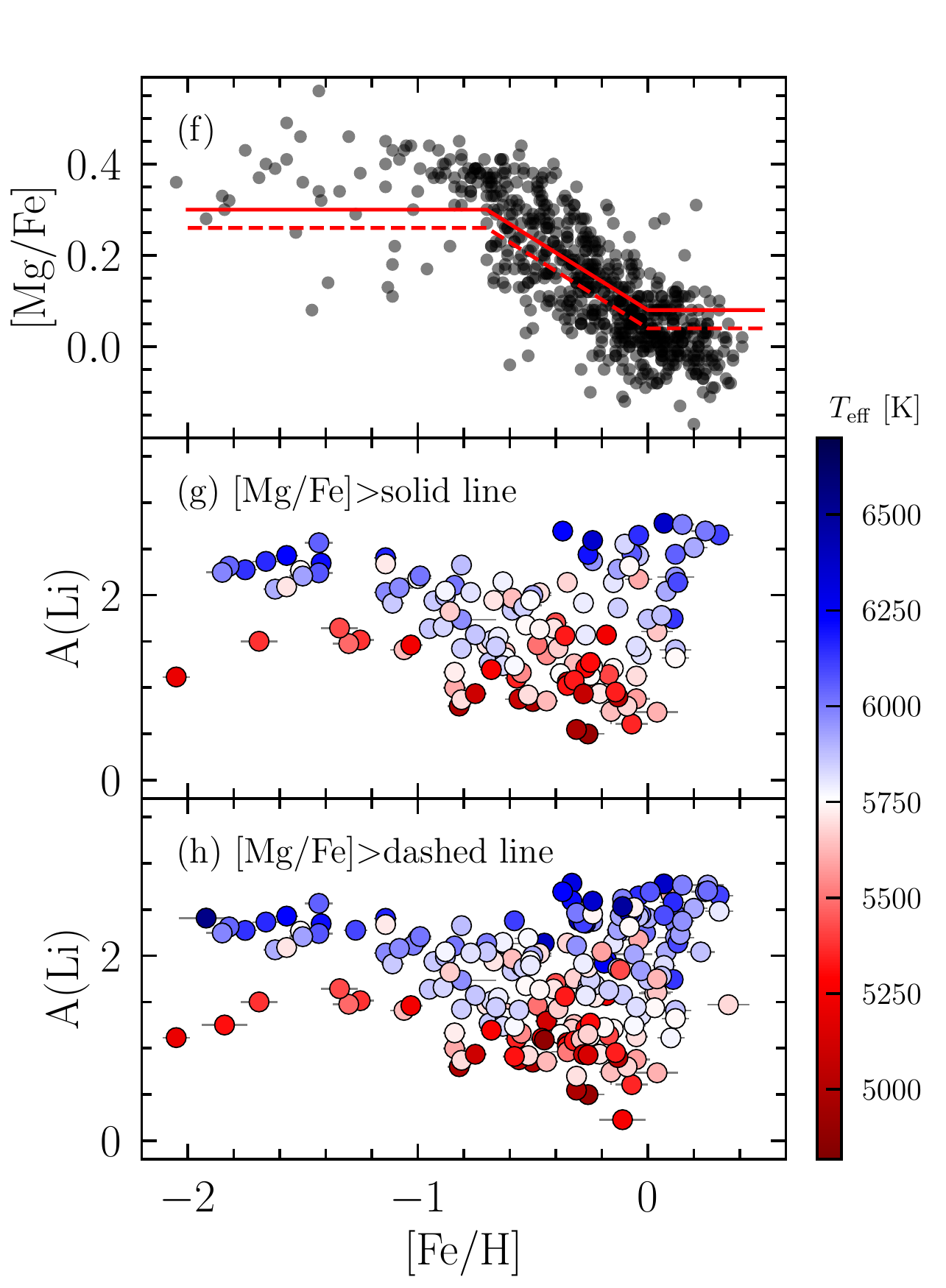}}
\caption{A(Li) versus [Fe/H] when selecting thick disk stars using kinematical or chemical criteria. (a) uses the kinematical criteria as defined in \cite{bensby2003} and selects stars that are at least 10 times more likely to be thick disk stars ($TD/D>10$), and (b) shows stars that are at least two times more likely to be thick disk stars ($TD/D>2$).  (c)-(h) uses $\alpha$-enhancement as a criterion to select thick disk stars. (d) shows the A(Li) trends when using [Si/Fe] as a proxy for $\alpha$-enhancement, requiring stars to have [Si/Fe] values above the solid red line in the [Si/Fe]-[Fe/H] plot shown in (c). In (e) we illustrate the effect of uncertainties in the abundance ratios by lowering the limit to the dashed red line in (c). The $\alpha$-enhancement criterion but using [Mg/Fe] instead is shown in (f)-(h). Upper limit Li abundances were not included in the plots.}
\label{fig:thickdisk_alpha}
\end{figure*}

\subsection{Li trends in the thin and thick disks}
\label{sec:thinthick}

It is not straightforward to define a criterion to select thick disk stars. As discussed in \cite{bensby2014}, the velocity distributions between the thin and the thick disk stellar populations are widely overlapping, making kinematically selected thin and thick disk samples significantly more contaminated than when using age as a criterion. In addition, a now commonly used criterion to identify thick disk stars is through the abundance ratios. We will therefore investigate how the Li trends in the thin and thick disks depend on how the samples are defined; by ages, by kinematics, or through chemistry.

\paragraph{\bf Age selection criteria:}

The thick disk has been found to be an older stellar population than the thin disk. The thick disk stars are generally older than about 8\,Gyr and the thin disk stars younger than about 8\,Gyr \citep[e.g.][]{haywood2013,bensby2014,kilic2017,silvaaguirre2018}. Whether or not this division is reflected by a hiatus in the star formation history of the Milky Way, and whether there is an overlap in age between the thin and thick disk stellar populations is still under investigation. It is however clear that age criteria to select thin and thick disk stellar samples results in cleaner and more well-defined abundance trends for the two disks \citep{bensby2014}. 
 
Figure~\ref{fig:thickdisk} shows the Li trends for our stars, separated into old ($>8$\,Gyr) and young ($<8$\,Gyr) subsamples, representing the thick and thin disks, respectively. For the old (thick disk) sample we see that the Li trend lies at the Spite plateau at metallicities below $\rm [Fe/H]\lesssim-1$. Then at metallicities above $\rm [Fe/H]\approx-1$ the Li trend shows a steady decrease, signalling that there is no significant Li enrichment, or that the depletion is larger than any production, during the later phases of the thick disk towards solar metallicities.   For the younger thin disk sample, shown in the bottom panel of Fig.~\ref{fig:thickdisk}, there is an increase in the Li trend, starting essentially at a level comparable to the Spite plateau at $\rm [Fe/H]\approx-0.8$ and then increasing towards higher metallicities. This signals a steady Li enrichment in the thin disk.

The clean and well-defined upper envelope on the thick disk A(Li) trend in Fig.~\ref{fig:thickdisk}a is encouraging, supporting the age criterion as a method to select thick disk stars, and that the A(Li) trend in the thick disk is truly declining with metallicity.

\paragraph{\bf Kinematical selection criteria:}

The `classical way' to select thick disk stars is to use kinematical criteria. In some sense this is what originally defined the thick disk, it is thicker, and its stars are moving on orbits that bring them farther away from the plane and/or closer to the Galactic centre. So they should have on average ``hotter'' kinematics than the stars of the thin disk. To do this separation we use the kinematical criteria defined in \cite{bensby2003}. Based on the $\ulsr$, $\vlsr$, and $\wlsr$ space velocities each star is given probabilities of belonging to either of the two disks, and $TD/D$ gives the ratio between these probabilities. $TD/D>1$ means that a star is more likely to be a thick disk star, and $TD/D<1$ means that it is more likely to be a thin disk star. 

Figures~\ref{fig:thickdisk_alpha}a and b show the $\rm A(Li)-[Fe/H]$ trends for candidate thick disk stars in our sample selected using kinematical criteria. The more stringent criterion where the likelihood of being a thick disk star is at least ten times larger than that of being a thin disk star ($TD/D>10$) shows a Li trend that is mostly flat. One or two stars may show a tendency of starting an increasing A(Li) trend around solar metallicities. For the $TD/D>2$ case, meaning less stringent criterion, but still at least two times more likely to be a thick disk star, there are more stars at higher metallicities that have high Li abundances that also increase with [Fe/H]. This is opposite to what is observed using the age defined thick disk sample in Fig.~\ref{fig:thickdisk} where the Li trend is decreasing with metallicity. 

The reason why the kinematical criteria is hard to use at high metallicities is because of the large overlap in the velocity distributions between the thin and the thick disk stellar populations. In combination with the facts that the stellar density of the thin disk population is about a factor of ten higher in the solar neighbourhood \citep[e.g.][]{blandhawthorn2016}, and that the thin disk population has a metallicity distribution that peaks around solar metallicities \cite[e.g.][]{casagrande2011}, while the thick disk metallicity distribution peaks around $\rm [Fe/H]\approx -0.6$ \citep[e.g.][]{gilmore1995,carollo2010}, will result in progressively more contaminated thick disk samples at higher metallicities. It should be noted that our stellar sample was defined to probe the extremes of the thin and thick disks, and hence contain many kinematically hot stars at higher metallicities. These are necessarily not thick disk stars but could rather come from the high-velocity tail of the thin disk. \cite{bensby2014} give more details on how the sample was selected.

\paragraph{\bf Chemical selection criteria:}

The thin and thick disks have experienced different chemical enrichment histories, which is manifested by the distinct and well-separated elemental abundance trends, in particular for the so called $\alpha$-elements \citep[e.g.][]{fuhrmann1998,bensby2003,reddy2006,adibekyan2012,bensby2014}. The thick disk can be traced all the way up to solar metallicities \citep[as shown by][]{bensby2007letter2}. This chemical distinction between the thin and thick disks is also seen in the large spectroscopic surveys such as Gaia-ESO \citep{recioblanco2014,mikolaitis2014,kordopatis2015} and GALAH \citep{duong2018}. Therefore chemical selection criteria could offer another option for selecting thick disk stars, when for instance stellar ages are not readily available. This was for instance done when inferring the short scale-length of the thick disk from a small sample of giant stars in the outer disk \citep{bensby2011letter}. However, as the abundance trends for the thin and thick disk converge towards solar metallicities, the separation becomes progressively more difficult at higher metallicities, and care should be taken if the abundances are not precise enough.

Figures~\ref{fig:thickdisk_alpha}$\rm c-h$ show the Li trends for candidate thick disk stars in our sample using chemical criteria. Here we use [Si/Fe] and [Mg/Fe] to represent the level of $\alpha$-enhancement. The scatter in [Si/Fe] is low and the thin and thick disk sequences are well separated, meaning that it is easy to define a separating line between the [Si/Fe] sequences in the two disks. The division is shown by the solid red line in Fig.~\ref{fig:thickdisk_alpha}c. The stars above this line are classified as thick disk stars and the plot below show the A(Li) vs. [Fe/H] trend for those stars. It is surprisingly clean and essentially mimics the Li trend for the age selected thick disk stars in Fig.~\ref{fig:thickdisk}. If the separation between the two disks would not have been as clear or if there were a larger scatter in the abundance ratios, Fig.~\ref{fig:thickdisk_alpha}e illustrates the outcome of lowering the separating line between the two disks by 0.04\,dex (dashed line in Fig.~\ref{fig:thickdisk_alpha}c). The picture now changes completely as the thick disk sample has been contaminated with thin disk stars. The Li trend is now instead increasing with metallicity. 

Not all abundance ratios are as well-defined as Si. The panel on the right-hand side of Fig.~\ref{fig:thickdisk_alpha} shows the case of Mg, an element that is the most often used element as a proxy for $\alpha$. The separation is now not as clear as for Si and it more difficult to define a dividing line between the two disks (solid red line in Fig.~\ref{fig:thickdisk_alpha}f). The resulting thick disk Li trend for those stars that lie above this line is shown in Fig.~\ref{fig:thickdisk_alpha}g. There is a steady increase in A(Li) towards higher [Fe/H] and the declining trend seen in the age-selected sample, or in the well-defined [Si/Fe]-selected sample, cannot be recognised. By lowering the line by 0.04\,dex as done for Si of course worsens the situation (Fig.~\ref{fig:thickdisk_alpha}h). It is clear that a chemical separation between the thin and thick disk becomes very difficult towards solar metallicities. 

For both kinematical and chemical selection criteria one can expect the thick disk samples to be progressively more contaminated by thin disk stars with increasing metallicity.

\section{Discussions}

\begin{figure*}
\resizebox{\hsize}{!}{
        \includegraphics{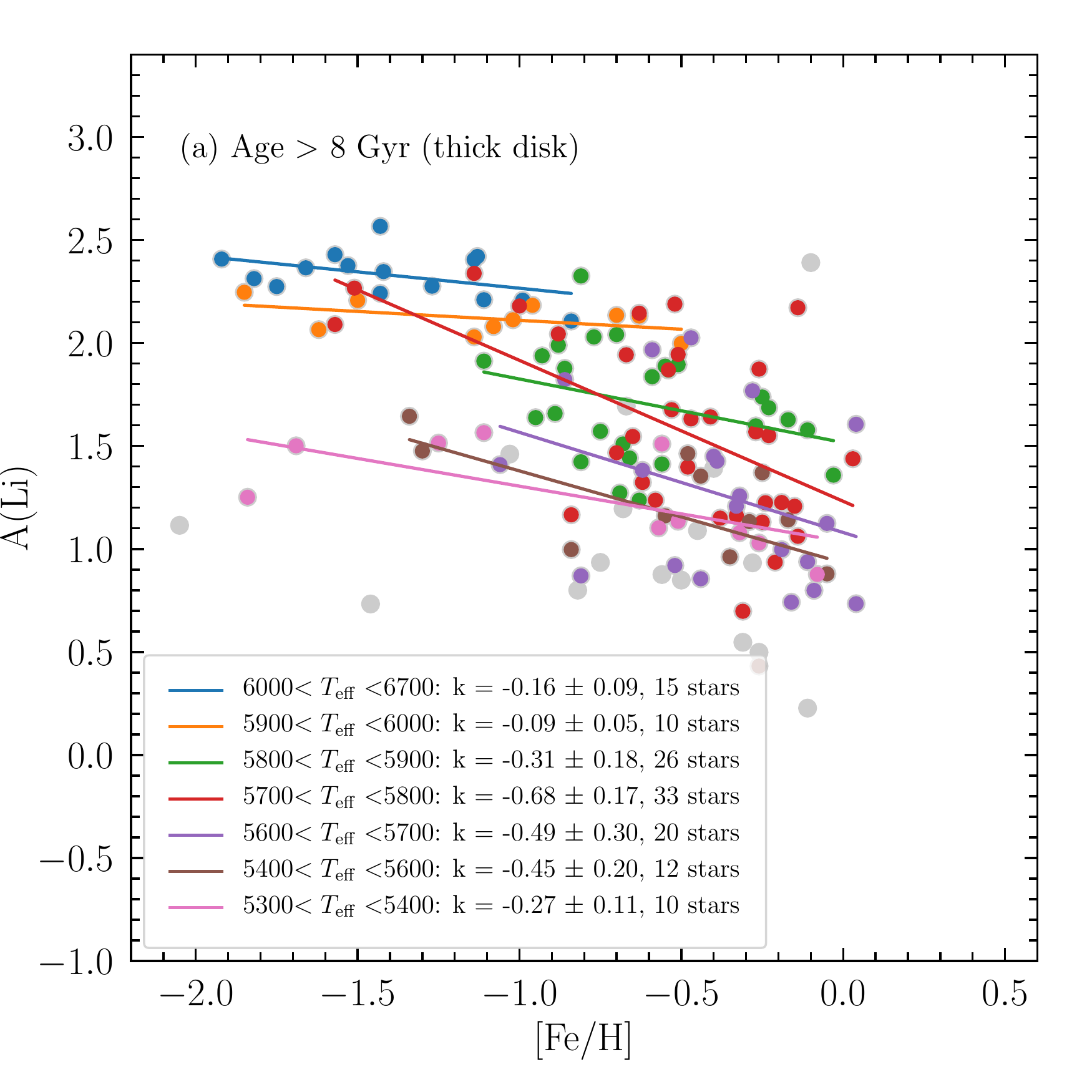}
        \includegraphics{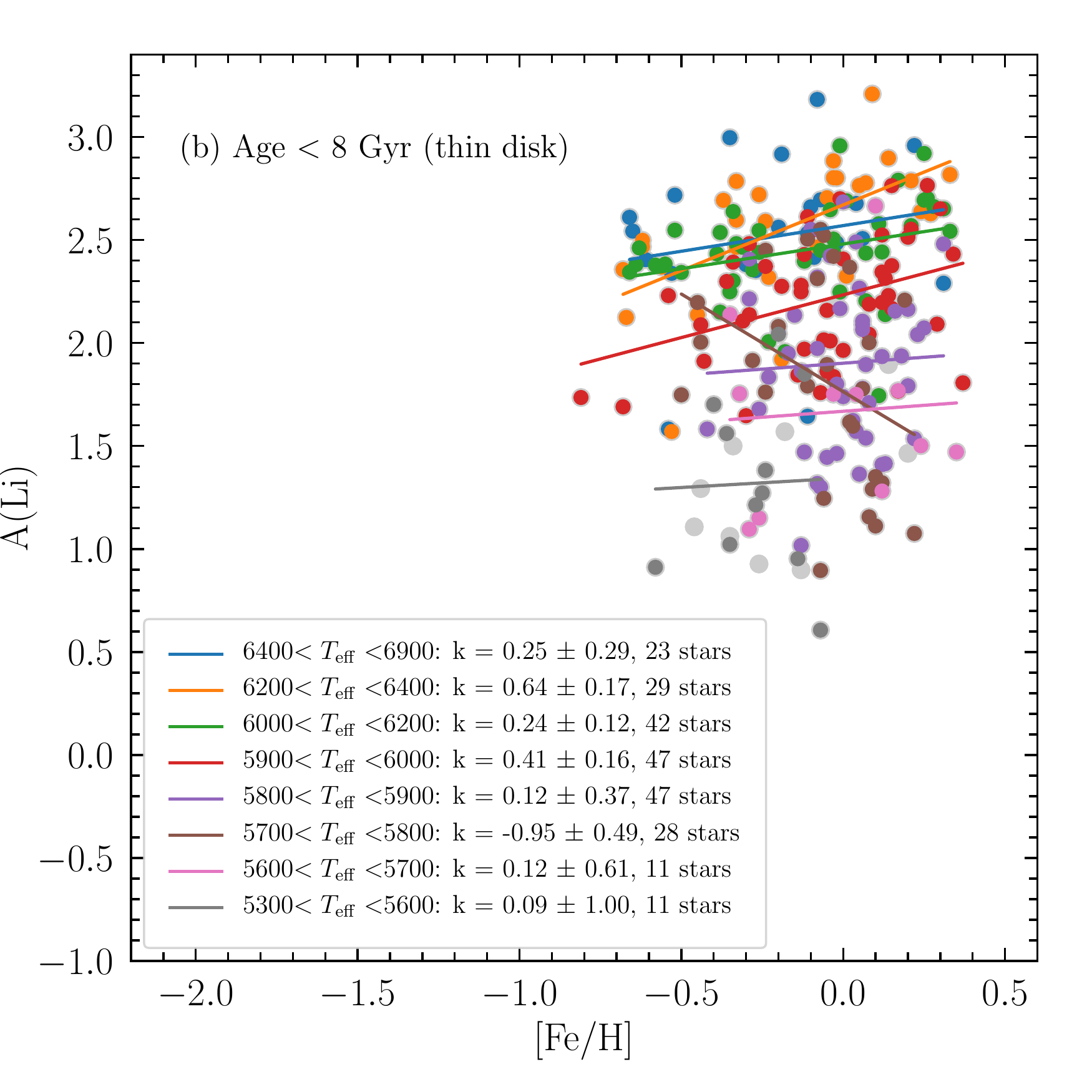}}
\caption{A(Li) versus [Fe/H] for the (young) thin disk and (old) thick disk, split into narrow $\teff$ slices. For each slice in $\teff$ a linear regression was performed, and the slopes (k) and associated uncertainties are given in the legends of the plots together with the $\teff$-intervals under consideration. Upper limit Li abundances were not included.
 }
\label{fig:thickdisk2}
\end{figure*}

Several recent studies have found that Li behaves differently in the thin and thick disks. \cite{ramirez2012} showed that the maximum Li abundance in thick disk stars appears to be nearly constant at a level very similar to what is seen in more metal-poor halo stars, and that any possible enrichment processes must have been erased from their atmospheres. The thin disk Li trend, on the other hand, showed an increase from the Spite plateau level and up to A(Li) values higher than three in the metal-rich thin disk. \cite{delgadomena2015} also found that the Li trends in the two disks differed, but that the thick disk Li trend decreased with metallicity rather than being flat.  More recently \cite{guiglion2016} used data from the AMBRE project and showed that the Li abundances increase with metallicity for thick disk stars, and even more for thin disk stars. This means that both disks should have experienced Li enrichment, but at different rates.  Also \cite{fu2018} using data from the Gaia-ESO survey \citep{gilmore2012} find that the thin and thick disks have different, but both still increasing, Li trends. Again a steeper increase in the thin disk, meaning more Li enrichment in the thin disk.

The studies above seem to agree that the maximum Li abundance increase in the thin disk, but disagree about how Li evolves in the thick disk, especially towards higher metallicities. Does it rise (as seen by \citealt{guiglion2016} and \citealt{fu2018}), is it flat (as seen by \cite{ramirez2012}), or does it decrease (as seen in this study and also possibly by \citealt{delgadomena2015}). What differs between the studies is how the selection of thick disk stellar samples were done. \cite{ramirez2012} used kinematical criteria (as defined in \citealt{bensby2003}), \cite{delgadomena2015} examined both kinematical criteria (as in \citealt{bensby2003}) and chemical criteria (as defined in \citealt{adibekyan2012}), while \cite{guiglion2016} and \cite{fu2018} used chemical criteria (defined by the level of $\rm [\alpha/Fe]$-enhancement, as described in \citealt{recioblanco2014}). 

Based on our investigation with the different selection criteria in Sect.~\ref{sec:thinthick} we believe that the rising Li trends for thick disk stars seen in the studies that use chemical selection criteria are due to contamination from the thin disk. As the Li in the thin disk shows a steady increase from $\rm [Fe/H]\approx-0.8$, this rise will to some degree be transferred to the (chemically defined) thick disk Li trends.

To further investigate the significances of the decreasing Li trend in the thick disk and the increasing Li trend in the thin disk Fig~\ref{fig:thickdisk2} shows again the A(Li)-[Fe/H] trends for the two subsamples divided into thin $\teff$-slices. Linear regression lines were fitted to the stars in each $\teff$-slice. For the (old) thick disk sample all $\teff$-slices have negative slopes, with uncertainties smaller than the values of the slopes. The slopes for the (young) thin disk sample $\teff$-slices have positive slopes, meaning increasing A(Li) trends with [Fe/H]. For both the thin and thick disk subsamples, colder $\teff$-slices show less significant slopes. These are temperature ranges where the stars have experienced a wider range of Li depletion. That A(Li) is decreasing with [Fe/H] for the thick disk, and increasing with [Fe/H] for the thin disk, for the hotter temperature intervals, appears to be statistically significant. Under the assumption that hot stars are equally resistant to Li depletion regardless of metallicity, we conclude that these findings have consequences for the Galactic production of Li.

A major result presented in this study is thus the clearly different and distinct A(Li)-[Fe/H] trends for the thin and thick disks (Fig.~\ref{fig:thickdisk}). The thin disk trend shows signatures of Li enrichment, while the thick disk trend shows signatures of depletion (Figs.~\ref{fig:thickdisk} and \ref{fig:thickdisk2}). This conclusion could also be seen in the run of A(Li) with stellar age in Fig.~\ref{fig:litrends}d that depicts that no major enrichment of Li occurred during the first four billion years in the evolution of the Milky Way stellar disk. As this is the era of the thick disk, this also means that no major Li production occurred then, or at least that any production was deprived by an even larger depletion.  

We have also shown, in agreement with many previous studies, that observed Li abundances in late-type stars show correlations with stellar effective temperature, surface gravity, metallicity, mass, and age, variables that are in turn connected to each other in a non-straightforward way (Figs.~\ref{fig:litrends}-\ref{fig:liagetwins}). These correlations reflect the complex physics of Li depletion and it is important that such stellar evolution phenomena are not biasing the interpretations about the Galactic evolution of Li. 

The impact of Li depletion is most clearly demonstrated in Fig.~\ref{fig:lizmass}, where mass (including implicitly also an age-dependence) and metallicity dependence are separated. The strong correlations seen in this plot can be naturally explained as a consequence of the larger convection zones in stars of higher metallicity and/or lower mass. However, since stars of lower metallicities have smaller masses at a given effective temperature, the demonstrated mass and metallicity dependence may also be interpreted as the manifestation of a simple dependence on effective temperature, explaining the tight Li-$\teff$ trend depicted in Fig.~\ref{fig:litrends}a.  At least, it appears that hot stars (between about 6000\,K to 6500\,K) are most robust against Li depletion regardless of metallicity.

\begin{figure*}
  \resizebox{\hsize}{!}{
          \includegraphics[viewport=0 0 504 504,clip]{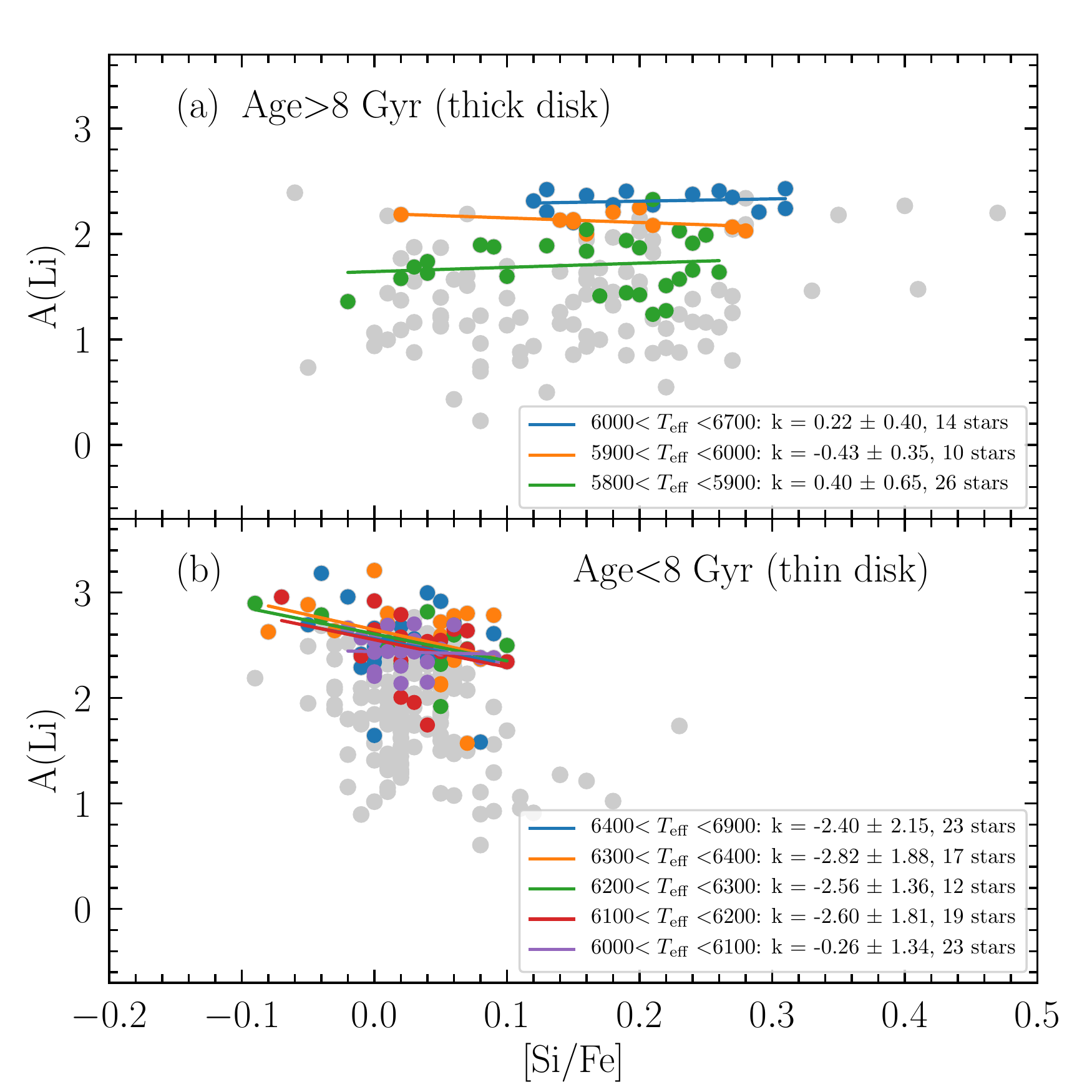}
          \includegraphics[viewport=0 0 504 504,clip]{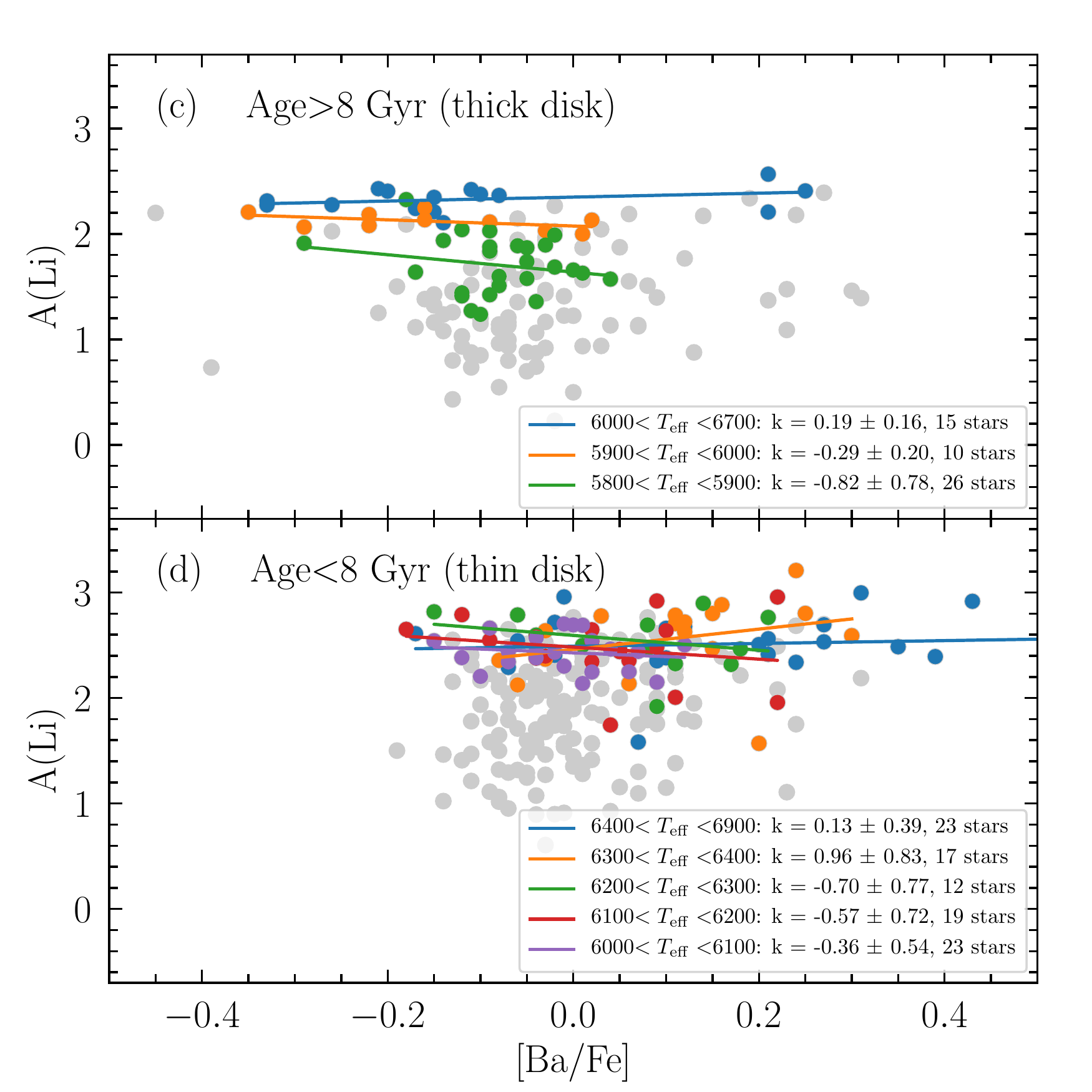}}
\caption{A(Li) versus [Si/Fe] (on the left-hand side) and versus [Ba/Fe] (on the right-hand side) for the (young) thin disk and (old) thick disk, split into $\teff$ slices. For each slice in $\teff$ a linear regression was performed, and the slopes (k) and associated uncertainties are given in the legends of the plots together with the $\teff$-intervals under consideration. Upper limit Li abundances were not included.
}
\label{fig:mgfe}
\end{figure*}

\cite{prantzos2017} compared the observed Li trends in the thin and thick disks with chemical evolution models and concluded that the production of Li in the Galactic disk(s) must have been due to a long-lived source, such as low-mass RGB stars, or else the Spite plateau value would have been elevated at lower metallicity. These evolve on time-scales longer than more massive AGB stars, and also longer than the time-scale of SNIa that are the main contributors to the enrichment of Fe. Although \cite{prantzos2017} based their conclusions on comparisons to the \cite{guiglion2016} data, that clearly shows indications of Li production in the thick disk, the same conclusion holds also here. We speculate that the assumptions about radial migration in the models of \cite{prantzos2017} could be better tuned so that the Li trends match the thick disk better. The thin disk is not very sensitive to assumptions about radial migration as it is a phenomenon that acts on long time-scales.

The recent study of \cite{fu2018} also investigated how A(Li) correlates with other abundance ratios. They found a $\rm A(Li) - [\alpha/Fe]$ anti-correlation, which was interpreted as an indication of more Li production during the Galactic thin disk phase. They also found a correlation between A(Li) and the $s-$process elements Y and Ba, interpreted as a connection between their nucleosynthesis sites being common, that is AGB stars. The decline in A(Li) at the highest metallicities could then be explained by a connection to a halt in the $s$-process, inferred from the apparent declines in the [Y/Fe] and [Ba/Fe] abundance trends seen at super-solar metallicities in the Gaia-ESO data. It should be noted here that the decline in [Y/Fe] at super-solar metallicities is not seen in the current sample (see Fig.~16 in \citealt{bensby2014}) and that a decline in [Ba/Fe] is only marginal (see again Fig.~16 in \citealt{bensby2014}). It should be noted that for stars hotter than about 6100\,K, Ba abundances could be significantly affected by NLTE effects, resulting in too high [Ba/Fe] abundance ratios. If those stars are included in the current sample, the [Ba/Fe] decline at super-solar metallicities becomes stronger.

In Fig~\ref{fig:mgfe} we show how A(Li) for the thin and thick disk subsamples vary with [Si/Fe] and [Ba/Fe]. Each plot also contain regression lines in different $\teff$-intervals, and we see that it is only for the thin disk subsample and only for [Si/Fe] that A(Li) shows any significant correlation. Following on the reasoning in \cite{fu2018} this could indicate that there was Li enrichment only during the thin disk phase (as A(Li) correlates with [Si/Fe] for the thin disk subsample), but that there is no evidence that Li should have been produced on the same time scale as Ba, that is AGB stars. As discussed above AGB stars evolve on a shorter timescales than RGB stars, and as there appears to have been no enrichment during the thick disk phase (the first 3-4\,Gyr of the evolution of the Milky Way disk), this is further strengthening the idea that long-lived RGB stars could have been one of the main producers and contributors of Galactic lithium.

Finally, it is important to connect our findings concerning stellar depletion and production in the Galactic disk to analogous studies for the Galactic halo. Recent observational advances have revealed that the Spite plateau of warm metal-poor halo stars, formerly thought to be flat and thin in nature, actually contains significant substructure. In particular, the extremely metal-poor end has been populated and while some stars are found on the Spite plateau \citep{bonifacio2018} many stars have Li abundances decreasing well below the Spite plateau and decreasing towards lower metallicities \citep{asplund2006,bonifacio2007,aoki2009,sbordone2010}. We stress that this is in fact opposite to what we have found for warm stars in the thick disk. Unless the behaviour can be explained by a change in efficiency of the Li depletion along the plateau, the extrapolation of halo Li abundance to zero metallicity would further aggravate the already substantial cosmological lithium problem. Arguments in favour of the depletion hypothesis have been raised by several studies, presenting observational evidence for a mass and metallicity dependence of Li abundance based on metal-poor field stars \citep{melendez2010li} and binaries with accurate mass determinations \citep{gonzalezhernandez2008,aoki2012}. The progressively lower masses encountered at lower metallicities could thus explain the lower Li abundances and larger scatter encountered in this regime. Our study corroborates these findings, indicating a strongly mass and metallicity dependent depletion over a large parameter range. We have demonstrated that stars of a given mass show a roughly linear anti-correlation between metallicity and Li abundance, qualitatively similar to the results for halo stars by \cite{nissen2012}. However, to draw stronger conclusions about a potential universal mechanism for Li depletion on the main sequence, one must conduct a homogeneous study of Li abundance in all Galactic populations. 

Furthermore, one must likely investigate the link to yet another parameter, in addition to mass, age and chemical composition, for example the rotation rate of the star. Another potential aspect to consider is the presence of planets, as suggested by \citet{israelian2009}. However, using a sample more restrictive in age and metallicity, \citet{baumann2010} later refuted the hypothesis that solar-like stars with planets have lower Li abundances than stars without planets. Also in this study we cannot see any statistically significant differences between stars with and without detected exoplanets.

\section{Summary}

We present NLTE corrected Li abundances for 420 stars from the sample of 714 nearby F and G dwarf stars by \cite{bensby2014}. For another 121 stars in the sample we present upper limits to their Li abundances. The stellar sample consists of stars belonging to both the thin and thick disks, allowing us to study the depletion and production of Li in the two disks separately. A small number (24 stars) of the stars that we could estimate the Li abundance for have also detected exoplanets.

Our main finding is that the evolution of Li is distinctly different in the Galactic thin and thick disks. While the Li abundance increases with metallicity in the thin disk, there is a steady decrease in the thick disk. This is in contrast to recent studies that have found the the Li abundances increase in both disks. We find that the reason for the apparent Li increase with metallicity in those studies is because they used chemical abundance criteria to define their thin and thick disk samples, leading to an increased contamination of thin disk stars into the thick disk sample with metallicity. As the thin disk Li trend increases with metallicity, this rise will be present as an apparent increase in the Li abundance with metallicity also for the thick disk. We have used a more robust age criteria to define thin and thick disk samples, and we find the thick disk Li trend to be clearly decreasing with metallicity. The conclusion is that there has been no significant Li production in the thick disk during the first few billion years in the history of the Galactic disk. This means that the production of Galactic lithium must have occurred in sources that evolve on time-scales that are similar or longer than those of the thick disk (a few billion years), which could be low-mass RGB stars.

We further confirm that the decrease in the Li abundance in the most metal-rich (thin disk) stars at super-solar metallicities. There is currently no good explanation for this decrease. Also, we see no difference in Li abundance between stars with and without detected exoplanets. The community appears to be divided in this question, and it seems as if the controversy lingers on. Our sample also appears to show an age-Li correlation for solar twin stars, and the gap in the $\teff$-A(Li) plane discovered by \cite{ramirez2012} is also void of stars in this study. 

It seems as if Li is an element that still poses many questions to be resolved. Maybe upcoming data releases from ongoing and future large spectroscopic surveys such as Gaia-ESO \citep{gilmore2012}, GALAH \citep{desilva2015}, and 4MOST \citep{dejong2016} can provide further pieces of the Galactic Li puzzle.

\begin{acknowledgement}

T.B. was supported by the project grant `The New Milky' from the Knut and Alice Wallenberg foundation. K.L. acknowledges funds from the Alexander von Humboldt Foundation in the framework of the Sofja Kovalevskaja Award endowed by the Federal Ministry of Education and Research as well as funds from the Swedish Research Council (Grant nr. 2015-00415\_3) and Marie Sklodowska Curie Actions (Cofund Project INCA 600398). We thank the referee Dr. Elisabetta Caffau, and also Elisa Delgado-Mena, for valuable comments on the submitted draft.  This research has made use of NASA's Astrophysics Data System Bibliographic Services.

\end{acknowledgement}
\bibliographystyle{aa}
\bibliography{referenser}

\end{document}